%% file: main.tex
\documentclass[twoside]{article}
\usepackage[accepted]{aistats2024}
\usepackage[round]{natbib}

\usepackage[T1]{fontenc}
\usepackage{graphicx}
\usepackage{hyperref}
\usepackage{color}
\usepackage[inline]{enumitem}       %inline enumeration
\usepackage{amsmath}                %math functions
\usepackage{booktabs}               %lines in tables
\usepackage[table,xcdraw,usenames,dvipsnames]{xcolor} %colors (also for tikz)
\usepackage{tikz}
\usepackage{standalone}
\usepackage{xspace}                 %correct spacing after newcommand macros
\usepackage{mathtools}              %for happy mathematicians (e.g., \coloneqq)
\usepackage{multirow}
\usepackage{pdfpages}               %include supplementary

\hypersetup{colorlinks, linkcolor={red!50!black}, citecolor={blue!50!black}}

\usetikzlibrary{shapes.geometric,fit,positioning,arrows.meta,external,calc}

\definecolor{luke_blue}{HTML}{1976d2}
\definecolor{luke_red}{HTML}{d32f2f}
\definecolor{luke_green}{HTML}{2e7d32}
\definecolor{my_yellow}{HTML}{FFCC67}

\newcommand{\diva}{\textsc{Diva}\xspace}
\newcommand{\falfa}{Falfa\xspace}
\newcommand{\abbrv}{\emph{\underline{D}etecting \underline{I}n\underline{V}isible \underline{A}ttacks}\xspace}

\newcommand{\tinytitle}[1]{\textbf{#1\quad}}
\newcommand{\eventiniertitle}[1]{\textit{#1:}}
\newtheorem{definition}{Definition}

\begin{document}

% If your paper is accepted and the title of your paper is very long,
% the style will print as headings an error message. Use the following
% command to supply a shorter title of your paper so that it can be
% used as headings.
%
%\runningtitle{I use this title instead because the last one was very long}

% If your paper is accepted and the number of authors is large, the
% style will print as headings an error message. Use the following
% command to supply a shorter version of the authors names so that
% they can be used as headings (for example, use only the surnames)
%
%\runningauthor{Surname 1, Surname 2, Surname 3, ...., Surname n}

\twocolumn[

\aistatstitle{Poison is Not Traceless: Fully-Agnostic Detection of Poisoning Attacks}

\aistatsauthor{ Xinglong Chang \And Katharina Dost \And Gillian Dobbie \And J\"org Wicker }

 \aistatsaddress{ The University of Auckland, Auckland, New Zealand\\
 xcha11@aucklanduni.ac.nz, \{katharina.dost, g.dobbie, j.wicker\}@auckland.ac.nz }
]

     %%%%%%%%%%%%%%%%%%%%%%%%%%%%%%%%%%%%%%%%%%%%%%%%%%%%%%%%%%%%%%%%%%%%%%%%%%%%%%% 
\begin{abstract}
The performance of machine learning models
depends on the quality of the underlying data. Malicious actors can
attack the model by poisoning the training data.
Current detectors are tied to either specific data types, models, or attacks, and therefore have limited applicability in real-world scenarios. 
This paper presents a novel fully-agnostic framework, \diva (Detecting InVisible Attacks),
that detects attacks solely relying on analyzing the potentially poisoned data set. 
\diva is based on the idea that poisoning attacks can be detected by comparing the
classifier's accuracy on poisoned and clean data and pre-trains a meta-learner using Complexity Measures to estimate the otherwise unknown accuracy on a hypothetical clean dataset.
The framework applies to generic poisoning attacks. For
evaluation purposes, in this paper, we test \diva on
label-flipping attacks. 

\end{abstract}

%%%%%%%%%%%%%%%%%%%%%%%%%%%%%%%%%%%%%%%%%%%%%%%%%%%%%%%%%%%%%%%%%%%%%%%%%%%%%%%
\section{Introduction}
% SUMMARY: A brief introduction on data poisoning attacks.
The performance of machine learning models relies strongly on the quality of the underlying data.
Bad actors can corrupt a model by injecting carefully crafted
malicious examples into the training data to significantly damage the
model's performance \citep{biggio2018wild,cina2022wild}.  
These {\em data poisoning attacks}
\citep{jia2021intrinsic}, are aimed at decreasing the model's
performance while keeping the data modifications to a minimum to avoid
detection.
Previous studies have shown that most machine-learning models are
vulnerable to such attacks
\citep{munoz2017towards,xiao2012adversarial,rosenfeld2020certified,zhang2020adversarial,aryal2022analysis}. 

This vulnerability is particularly problematic with applications that integrate user-submitted data into their machine learning pipeline \citep{zhang2021data}, due to the ease of access to the training data. 
In these cases, confirming that the training data is not poisoned before running other data or model selection processes is critical to avoid deploying and using flawed models. 
If a dataset were found to be poisoned, it could be excluded from the pipeline, or sanitization techniques can be employed until it can be confirmed as clean.

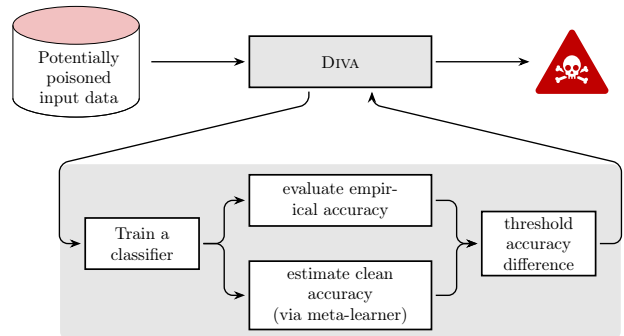
\begin{figure}
    \centering
    \resizebox{\columnwidth}{!}{\input{tikz/diva.tikz}}
    \caption{
        To decide whether an input dataset is poisoned, \diva trains a classifier, evaluates its empirical accuracy on the input data and simultaneously estimates its accuracy on clean data. If their difference exceeds a threshold, the dataset is flagged as poisoned.
        % \diva pre-trains a meta-learner and extracts {\em Complexity Measures} (C-Measures) from the potentially poisoned data to estimate the clean accuracy.
        % If the discrepancy exceeds a threshold, the dataset is flagged as poisoned.
    }
    \label{fig.diva}
\end{figure}

% motivation why we need agnostic
We conclude that a detector with \textit{agnostic} capabilities is essential. In particular, the detector should be 
\begin{enumerate*}
    \item \textit{data- and datatype-agnostic}, i.e., neither constrained by dataset dimensions or distributions nor by the data type to transfer concepts and knowledge across domains and allow for a better generalization of the detector \citep{vanschoren2019meta};
    \item \textit{attack-agnostic}, i.e., not tied to specific attacks, since we cannot expect to know in advance which attack an adversary will use; and 
    \item \textit{model-agnostic}, i.e., independent of specific machine learning models. The latter is essential to validate a dataset (once) before using it for multiple models that might be used in the future ensuring long-term efficiency.
\end{enumerate*}
Lastly, a detector cannot rely on the availability of a clean validation set, as all data should be considered potentially poisoned.
% Such an approach is indispensable as it facilitates broad applicability and long-term efficiency. This notion of agnosticism extends beyond current techniques that focus on specific aspects, as discussed in the prior paragraph, and offers a promising avenue for addressing the complex challenge of detecting strong poisoning attacks.

% notes
% data-agnostic -> no constraints regarding dataset dimensions, distributions
% datatype-agnostic -> we work for all types of data, incl. images and sound and tabular / we don't limit our defence / detection to specific types of data -> important in multi-modal settings
% attack-agnostic -> we don't know what the attacker will use
% model-agnostic -> independent of model works best for our specific use case; can validate the dataset before choosing one or multiple models that might be used in the future; amortized cost over time
% [task-agnostic -> classification and regression]

Defenses against poisoning attacks commonly focus on:
\begin{enumerate*}
    \item \textit{Point-wise sanitization}, which filters out poisoning
    examples and outliers \citep{paudice2018label}. This technique is often model-, data-, datatype-, or attack-specific and 
    typically requires a clean validation set
    \citep{steinhardt2017certified}. However, clean data is often
    unavailable when the validation and training sets come from the same
    source. 
    \item \textit{Certified robustness}, which provides a certificate
      guaranteeing the classifier's resilience against attacks up to a
      certain threshold \citep{li2019certified}, but 
      is model-specific.
      %has limited       utility when a large portion of the data is poisoned
      \citep{rosenfeld2020certified,zhu2022adversarially}. 
\end{enumerate*}
% Thus, detecting strong poisoning attacks remains a complex task
% with no straightforward automated solution. 
None of these defenses are sufficient in a real-world setting.

%Rather than limiting itself to a model or individual data points, a detector should detect poisoning on a dataset level to be prepared for the diverse threats training and poisoning scenarios pose in a real-world setting.

% Defences against poisoning attacks commonly focus on point-wise sanitization, which filters out poisoning examples and outliers \citep{paudice2018label}, or robust learning, which enhances the classifier's resilience against attacks \citep{rosenfeld2020certified,li2019certified,lecuyer2019certified}.
% Pointwise sanitization typically requires a clean validation set \citep{steinhardt2017certified}, which is not available for many real-world applications -- usually, only one dataset is available, and there is no indication of which parts are poisoned and which are clean. 
% Robust learning, on the other hand, can improve model robustness by inducing adversarial examples into the training process, but it does not test the dataset itself.
% Thus, poisoning detection remains a complex task with no straightforward automated solution.
% However, both approaches typically require a clean validation set \citep{steinhardt2017certified}, which is not available for many real-w
% orld applications -- usually, only one dataset is available, and there is no indication of which parts are poisoned and which are clean. Thus, poisoning detection remains a complex task with no straightforward automated solution.

Independent of the concrete attack scenario, the main goal of an attacker is to degrade the classifier's performance at prediction time \citep{biggio2018wild,demontis2019adversarial}, which is a general observation that does not tie to any specific dataset, -type, model, or attack.
If we obtain a poisoned dataset, we can expect to observe a lower empirical performance than what a classifier is theoretically capable of on a clean dataset.
Conversely, a clean dataset should exhibit an empirical accuracy that matches its clean accuracy.
Therefore, we hypothesize that comparing both performances is key to distinguishing between clean and poisoned datasets. 
However, we have no means to evaluate the clean accuracy directly.

Building on this observation, we propose \diva (\abbrv), a novel framework that
detects poisoning attacks by evaluating the discrepancy between the
classifier's estimated accuracy on hypothetical clean data and the empirical accuracy on the observed data (Figure \ref{fig.diva}).
% To estimate the clean accuracy, \diva uses a meta-learner \citep{vanschoren2019meta}.
% Unlike existing methods, \diva does not require a clean dataset for detection and takes into account the entire dataset rather than individual examples. 
Given only the potentially poisoned dataset, \diva quantifies the strength of the poisoning effect on a model's performance, rather than reflecting the number of poisoned examples.

To estimate the clean accuracy, \diva pre-trains a meta-learner \citep{vanschoren2019meta} using {\em Complexity Measures} (C-Measures) \citep{ho2002complexity,lorena2019complex} to capture the distinct patterns of poisoned data using meta-data extracted from other datasets. 
First, this approach is data-, datatype- and model-agnostic since we employ a huge meta-database of models and datasets, making it applicable to diverse datasets. Second, \diva is attack-agnostic as all attacks share the same goal -- maximizing the difference in performance between poisoned and clean models -- and \diva uses only this assumption in the detection. 
%we show that the detection of attacks transfers from the attacks in the meta-database to new previously unseen attacks.
% C-Measures closely correlate to a classifier's performance; we transfer this approach to detecting poisoning attacks.
% This allows \diva to estimate the expected accuracy of a model without the need for a clean validation set.
%
% A typical use case for \diva is to identify malicious clients.
% Applications, such as malware detection \citep{aryal2022analysis} and federated learning \citep{li2021detection}, rely on data from clients, which cannot be fully trusted. 
% % The negative impact of deploying poisoned models is significant. 
% \diva addresses this by providing a way to detect poisoned training datasets, offering a pre-processing sanity check that can lead to the identification of malicious clients. 
%
% A typical case for \diva is in the domain of machine learning service providers. These providers rely on data from clients, which cannot be trusted to be free of poisoning by malicious actors, and clean sets are not necessarily available. However, the effect of deploying models that are not performing well is significant. \diva addresses this by providing a way to test data for poisoning offering a pre-processing sanity check. 
%

We summarize our contributions as follows:
\begin{enumerate*}
    \item We define the problem of fully-agnostic detection of poisoning attacks.
    \item We are the first to identify the relationship between
      C-Measures and data poisoning attacks and use this
      relationship to detect poisoning attacks at a dataset
      level, even when a clean validation set is not available. 
    \item We propose \diva (\abbrv), a fully-agnostic
      meta-learning framework that uses C-Measures to quantify data poisoning
      attacks. Experiments show that \diva can reliably detect unknown
      attacks when presented with only poisoned data, particularly against
      extremely poisoned data where existing defenses tend to fail. 
    % \item To enable efficient meta-learning on sufficiently large
    %   data, we propose a novel label-poisoning attack, \falfa ({\em
    %     Fast Adversarial Label Flipping Attack}), that provides a
    %   substantial speed-up over existing attacks. 
\end{enumerate*}

 The paper is structured as follows: Section \ref{sec.prob} describes the problem setting, while Section \ref{sec.diva} and Section \ref{sec.exp} explain and evaluate the \diva framework. Section \ref{sec.relatedwork} covers related work, and Section \ref{sec.conclusion} concludes the paper.

\section{Scope and Problem Setting}
\label{sec.prob}

Data poisoning attacks occur when adversaries take control of the
training data. By manipulating the training data, the adversaries aim
to reduce the model's performance at inference time. 

% What are we doing and why
% In this paper, we introduce the novel black-box detection framework \diva for poisoned datasets, particularly for detecting label poisoning attacks. Unlike previous work, \diva does not require a non-poisoned ground truth and considers the whole dataset instead of single instances. 

% What is input, what is output
% \diva takes as input a single potentially poisoned dataset and ultimately returns a score that estimates if it is poisoned or not. This differs from previous approaches where an additional clean validation set was needed, which is hard to obtain in real-world scenarios. We also focus on static non-changing datasets as there is a lack of reliable poisoning detectors.

\tinytitle{Problem Setting}
%
% context
In a poisoning attack, the attacker can directly manipulate the training data and interact with the training process of a model \citep{biggio2018wild}. Based on this information, the attacker creates a poisoned training set, on which a poisoned model is trained by modifying or injecting examples. To increase the test error, the attacker's goal is to maximize the loss on the clean test set.

The success of an attack and therefore the need for a detector depends on a number of factors. One factor is the difficulty of the dataset in terms of the learning task. If a dataset is easy to classify and a model achieves good accuracy, a successful poisoning attack will lower the accuracy on the test set. On the other hand, on a dataset where it is hard to train a well-performing model, poisoning methods cannot lower the accuracy much further, as it is low already. However, as the goal of the attack is to maximize the difference in the performance on the test set compared to the training set, it will actually try to raise the accuracy on the training set, overfitting the model.
Either way, we can expect a high poisoned (training) accuracy together with a low clean (test) accuracy supporting \diva's approach to estimate and compare both.
%
% \begin{figure}[t]
%     \centering
%     \includegraphics[width=\columnwidth]{images/fake_acc.pdf}
%     \caption{The theoretical performance degradation from poisoning
%       attacks at various rates for easy and hard datasets. } 
%     \label{fig.fake}
% \end{figure}
%

Note that this is the most interesting case for detectors as summarized in Table \ref{tab:attack-matrix}. In cases where a high training accuracy is achieved, the performance of the model is at stake. If the training accuracy is already low, the attacker can only achieve an obvious attack (overfit model), or a failed attack (raising test accuracy).
% \diva targets cases where a high training accuracy is achieved and the performance of the model is at stake. In case the training accuracy is already low, the attacker can only achieve an obvious attack (overfit model with low training or test accuracy), or a failed attack (raising the test accuracy) (see Table \ref{tab:attack-matrix}). 

\begin{table}[t]
    \centering
    \caption{In a poisoning attack, the threat is negligible if
      training accuracy is lower than expected (yellow) as the models
      will not be used with a low performance. If 
      test accuracy is high, the attack was not successful and no
      detection is required. \diva operates in the
      red region, where a successful attack yields high training
      accuracy and low testing accuracy.}
    \scriptsize
    \setlength{\tabcolsep}{1em}         % adjust horizontal padding here
    \renewcommand{\arraystretch}{1.4}   % adjust vertical padding here
    \begin{tabular}{>{\centering\arraybackslash}m{1.8cm}|>{\centering\arraybackslash}m{1.8cm}>{\centering\arraybackslash}m{1.8cm}}
        & \textbf{Low Training} Accuracy & \textbf{High Training} Accuracy \\\hline
        \textbf{Low Test} Accuracy
            & {\cellcolor{my_yellow!30}Obvious Attack}
            & {\cellcolor{luke_red!30}Successful Attack} \\
        \textbf{High Test} Accuracy
            & {\cellcolor{my_yellow!30}Failed Attack}
            & {\cellcolor{luke_green!30}No/Failed Attack} \\
    \end{tabular}
    \label{tab:attack-matrix}
\end{table}
% \begin{table}[t]
%     \centering
%     \caption{In a label poisoning attack, the threat is obvious if training accuracy is lower than expected (yellow). \diva operates in the red region, where a successful attack yields high training accuracy and low testing accuracy.}
%     \scriptsize
%     \setlength{\tabcolsep}{1em}         % adjust horizontal padding here
%     \renewcommand{\arraystretch}{1.4}   % adjust vertical padding here
%     \begin{tabular}{c|cc}
%         & \textbf{Low Training} Accuracy & \textbf{High Training} Accuracy \\\hline
%         \textbf{Low Test} Accuracy
%             & {\cellcolor{my_yellow!30}Obvious Attack}
%             & {\cellcolor{luke_red!30}Successful Attack} \\
%         \textbf{High Test} Accuracy
%             & {\cellcolor{my_yellow!30}Failed Attack}
%             & {\cellcolor{luke_green!30}No Attack} \\
%     \end{tabular}
%     \label{tab:attack-matrix}
% \end{table}

\tinytitle{Scope}
The concept is designed for general poisoning attacks and it is fully-agnostic:

\begin{definition}[Fully-agnostic]
    An adversarial defense is \emph{fully-agnostic} if it does not impose any restrictions on 
    \begin{enumerate*}[label=(\roman*)]
        \item the specific characteristics, dimensions, or distributions of a dataset (\emph{data-agnostic}), 
        \item the dataset type, e.g., image, tabular, or graph data (\emph{datatype-agnostic}), 
        \item the model type or architecture (\emph{model-agnostic}), or
        \item the attack or attack scenario under which an adversary operates (\emph{attack-agnostic}).
    \end{enumerate*}
\end{definition}

For
evaluation purposes, in this paper, we implement and test \diva on
label-flipping attacks on binary classification as these attacks are particularly accessible to
attackers and can pose a serious threat to the performance of a model. 
In a \emph{label-flipping attack}, the attacker deliberately mislabels the
training data points to deceive the model into making incorrect
predictions \citep{munoz2017towards}.
In a multi-class setting, the user can include multi-class datasets into the meta-database or proceed in a one-versus-rest setting.
Future experiments are required to confirm similar behavior on
general poisoning attacks.

% So far, we present the algorithm in a binary classification setting.
% On a multi-class dataset, the attacker can target one class by applying the {\em one-versus-rest} (OVR) technique, merging other classes except the targeted one, resulting in the model misclassifying other classes as the targeted class.
% Similarly, we can use the OVR strategy when computing C-Measures on multi-class data.
% A dataset is poisoned if any pairwise measure exceeds the threshold.

\section{\diva: Detecting Invisible Attacks}
\label{sec.diva}

We discussed that comparing the empirical with the theoretical clean
accuracy can reveal poisoning (see Section \ref{sec.prob}). % (Figure \ref{fig.fake}).
To apply this concept in practice, \diva measures the empirical
accuracy and estimates the theoretical clean accuracy using
meta-learning. Instead of learning models from datasets directly,
meta-learning extracts information from models and datasets. This
forms a dataset and is used to ``learn to learn''. 
In our case, we use a meta-learner to estimate, given a potentially poisoned dataset, the hypothetical performance of a model if the dataset was clean.
%In our case, we predict the potential performance of a model on a hypothetical clean test set based on a potentially poisoned dataset. 

\begin{figure}[]
    \centering
    \resizebox{0.9\columnwidth}{!}{\input{tikz/metabase_vertical.tikz}}
    \caption[]{
        \diva pre-trains a meta-learner as follows:
        \begin{enumerate*}
            \item Collect multiple datasets; one of them is the input dataset here.
            \item Split each dataset into training and testing sets and generate multiple poisoned datasets from the training set.
            \item Create a meta-database by extracting C-Measures from the clean and poisoned datasets as features and using the classifiers' accuracies as targets.
            \item Train the meta-learner.
        \end{enumerate*}
    }
    \label{fig.metadatabase}
\end{figure}
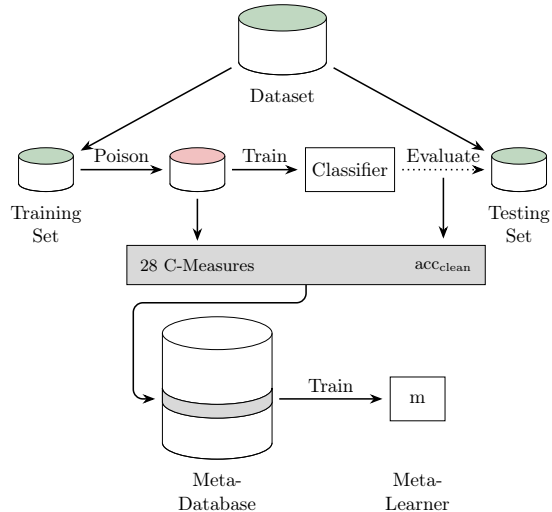

\diva pre-trains a meta-learner on
meta-data gathered from other datasets (Figure \ref{fig.metadatabase}),
applies it to estimate the clean accuracy (Figure~\ref{fig.diva}), and
finally compares the estimated accuracy to the observed accuracy on a
potentially poisoned dataset. 
This section explains \diva's components in detail.
% However, as testing accuracy cannot be measured without clean samples, \diva pre-trains an estimator on metadata with both clean and poisoned data available (Figure~\ref{fig.metadatabase}), and then applies it to the potentially poisoned dataset (Figure~\ref{fig.diva}).
% \diva is a {\em black-box} dataset-level detector, as the meta-learner is pre-trained on other datasets and has never seen the potentially poisoned dataset.

\tinytitle{Collecting Datasets for the Meta-Learner}
% In a given machine learning task, to estimate the clean accuracy of a potentially poisoned training set (denoted as $\mathcal{D}^{*}_\text{tr}$) using \diva, we need to populate the meta-database with clean and poisoned data first.
% To allow the meta-learner to learn the relationship between the extracted feature representations using C-Measures (introduced in Section~\ref{sec.cmeasures}) and the labeled poisoning attacks, we collect multiple datasets $G^{(i)}$ for the meta-learner (Figure~\ref{fig.metadatabase}).
%
The first step of \diva is to collect multiple datasets; each yields one instance for the meta-database on which a meta-learner can train.
To improve the generalizability of the trained meta-learner, we create a diverse database using the following methods:
\begin{enumerate*}
    \item\textbf{Synthetic data generation:} We generate multiple synthetic datasets with similar characteristics, such as the distribution of features and classes.
    \item\textbf{Real-world data:} We use standard datasets that share
      a similar task. \diva has no constraints on neither feature nor
      class distribution.%Since \diva does not require datasets to be in the same feature space, we can use open-source data.% that shares a similar task.
\end{enumerate*}
The important part is to extract meta-data from these datasets, so \diva can learn the relationship between the meta-data and the classifiers' accuracies.
The content of the data is less important to \diva, as long as they share a similar task, e.g., binary classification.
We show in Section \ref{sec.exp} that these datasets do not necessarily have to share the same distribution as the targeted training set.

\tinytitle{Generating Clean and Poisoned Pairs}
\label{sec.meta}
After obtaining the datasets, we split each into training and test sets (see Figure \ref{fig.metadatabase}).
We then use the training set to generate multiple poisoned datasets.
% , denoted as $G^{(i)}_{\text{p}j}$.
% We call the attack that generates poisoned data a {\em poison generator}.
For each {\em poison generator} (an attack that generates poisoned data), we apply a full spectrum of poisoning rates to obtain diverse inputs for the meta-learner.
We additionally perturb the training set using {\em Stochastic Label Noise} (SLN) and also proceed with the clean dataset as it is, so the meta-learner can learn to distinguish between poisoned and clean data.
% Ideally, each generator uses a distinct method to create poisoning examples, such as random label flipping, and ALFA ({\em Adversarial Label Flipping Attack})\citep{xiao2012adversarial}.

\diva does not require many poison generators as long as the attack is
{\em error-generic} \citep{biggio2018wild} as previous research proved
the transferability of poisoning attacks \citep{demontis2019adversarial}.
This is essential as new attacks are proposed regularly, and we need
to ensure the meta-learner can generalize to new attacks.
In addition, the attack we use should be {\em effective} so we address high-threat situations (as in Table \ref{tab:attack-matrix}).
Lastly, we need a {\em fast} attack as meta-learning requires repeated attacks on multiple datasets to build a meta-database and can be computationally costly.
% \diva does not require many poison generators as long as the attack has the following properties:
% \begin{enumerate*}
%     \item The classification error should be positively correlated to the poisoning rate, 
%     \item the computation time should be invariant to the poisoning rate, and
%     % to enable a wide range of poisoned datasets for the meta-database.
%     \item the attack should be error-generic to make sure the meta-learner can generalize to new attacks, which are proposed regularly, as every error-generic attack shares the same goal.
% \end{enumerate*}
%New attacks are proposed regularly and the success rates for every attack may vary, but every error-generic poisoning attack shares the same goal enabling the meta-learner to generalize to new attacks.
%
Most existing label poisoning attack \citep{xiao2012adversarial,chan2018data,paudice2018label,zhang2020adversarial,rosenfeld2020certified} fulfils all three properties. 
We implement a recently proposed poisoning attack, \falfa
%({\em Fast Adversarial Label Flipping Attack})
\citep{chang2024fast}, which is directly derived from the attacker's goal and transforms it into a linear programming task achieving a substantial speed-up over existing attacks, which makes it suitable for the creation of \diva's meta-database. 

%\tinytitle{Fast Adversarial Label Flipping Attack}

% Equation~\ref{eq1} is the general form of a poisoning attack.
% In a label poisoning attack, the attacker's capability is restricted to modifying labels.
%In a binary classification task, given $\ytr:=\{y_i\}^n_{i=1}$ and the percentage $\epsilon$ of examples  the adversary can modify,
%In a binary classification task, given $\ytr:=\{y_i\}^n_{i=1}$, and $\epsilon$ is the percentage of examples that the adversary can modify.
%the label poisoning attack wishes to find the optimal solution of $\ypo:=\{y'_i\}^n_{i=1}$ to maximize the difference between $f$ and $f'$:
%
% \begin{equation}
%      \footnotesize
%      \min_{\Dpo} \ell(\Dpo, f') - \ell(\Dte, f')
%     \quad\Leftrightarrow\quad \left\{
%     \begin{aligned}
%         \min_{\ypo} \quad   & \ell(f'(\Xtr), \ypo) - \ell(f(\Xtr), \ypo)           \\[-0.5em]
%         \textrm{s.t.} \quad & \sum_{i=1}^{n} |y'_i - y_i| \leq n \epsilon,         \\[-0.3em]
%         \quad               & y'_i \in \{0, 1\}  \; \text{for } i = 1,  \ldots, n.
%     \end{aligned} \right\}
%     \label{eq2}
% \end{equation}
%
%\begin{equation}
%     \footnotesize
%     \min_{\Dpo} \ell(\Dpo, f') - \ell(\Dte, f').
%     \label{eq1}
%\end{equation}
%
%

% \falfa is not limited to neural networks.
% It applies to any classifier that uses a Cross Entropy Loss.
% In fact, all components of the \diva framework are designed to work on a wide range of classifiers.

\tinytitle{Creating a Meta-Database}
For the meta-database, we collected a large number of datasets, both synthetic and real-world. On each dataset, we apply \falfa to have a clean, non-poisoned version, and poisoned version. Since attacks transfer \citep{demontis2019adversarial}, we do not need to include additional attacks. Our goal is to train a model based on this database that is able to predict the test accuracy of a new potentially poisoned dataset. 
We train classifiers on both clean and poisoned sets and obtain their accuracies on the clean testing set.

Next, we extract descriptive features from both clean and poisoned datasets as inputs for a meta-learner.
% These features serve as inputs for the meta-database. 
Precisely selecting these features is crucial as the learner's predictions depend on them, and they must accurately capture the classification task's complexity while avoiding deception from poisoning examples.
See Figure \ref{fig.metadatabase} for an overview.

\eventiniertitle{Complexity Measures}
C-Measures \citep{lorena2019complex,branchaud2019spectral} quantify the complexity of a supervised classification task \citep{ho2002complexity} by measuring the overlap between features, class separability, geometry, structural representation of a dataset, etc.
All these measures closely relate to the classifier's performance \citep{lorena2019complex,brun2018framework,leyva2014set}.
C-Measures are excellent candidates for \diva for two reasons:

First, prior studies show that a subset of C-Measures has a strong correlation to stochastic label noise (SLN) and C-Measures can be used to estimate the rate of noisy labels in a dataset \citep{garcia2015effect,saez2013predicting,wu2021triple}.
% Label noise can be considered as a suboptimal poisoning attack \citep{wu2021triple}.
In our case, by applying SLN as a poison generator, we extract meta-data from clean, noisy, and poisoned data, so the meta-learner trained on these meta-data can learn the difference between noisy and poisoned data.
% SUMMARY: Adversarial transferability

Second, the success rate of attacks falls after the adversarial examples transfer to other classifiers \citep{papernot2016transferability}.
C-Measures are based on linear discriminant analysis, principal component analysis, linear {\em Support Vector Machine} (SVM), {\em k-Nearest-Neighbors} (kNN) classifier, decision tree, and graph models \citep{lorena2019complex}.
Prior studies showed that adversarial examples can transfer between some but not all classifiers \citep{demontis2019adversarial,papernot2016transferability}, so an inconsistency between different classifiers can be an indicator of poisoning.

% SUMMARY: What C-Measures are we using?
% C-Measures rely on training basic classifiers on the training data, and the mean value from a 5-fold cross-validation is reported. 
\diva uses a C-Measure set that contains 22 normalized attributes and 5 additional attributes (i.e., standard deviations) as meta-data, allowing for comparisons between datasets of different sizes and dimensions. 
Find the full set of measures in the supplementary material.
Finally, we assemble the meta-database by translating each dataset into one row in the form 
$\text{row} = [\text{C-Measures}; \ \text{acc}_\text{clean}],$
%\[\text{row} = [\text{C-Measures}; \ \text{acc}_\text{clean}],\]
where $\text{acc}_\text{clean}$ is the clean test accuracy of the trained classifier.

Note that a performance drop in the clean (testing) accuracy can also be caused by overfitting, e.g., due to noise \citep{han2018co}. This scenario is similar to an obvious attack (see Table~\ref{tab:attack-matrix}).
To prevent \diva from reporting false positives, we put two methods in place:
\begin{enumerate*}
    % \item We use {\em Cross-Validation} (CV) to ensure classifiers have optimal parameters to prevent overfitting.
    \item We apply techniques that prevent overfitting, e.g., {\em Cross-Validation} (CV) and normalization, when training the individual classifiers corresponding to the datasets in the meta-database. 
    \item We use SLN as a poison generator for the datasets in the meta-database \citep{chen2020noise}, so \diva learns to distinguish the patterns from clean, noisy, and poisoned data.
\end{enumerate*}

\tinytitle{Meta-Learner and Threshold}
\label{sec.estimator}
We train a meta-learner on the $[\text{C-Measures}; \ \text{acc}_\text{clean}]$ data collected in the meta-database that takes the C-Measures as an input and learns to predict the clean accuracy.
%Using the meta-database, we set up the machine learning task with C-Measures as attributes and clean testing accuracies as the targets (see Figure \ref{fig.metadatabase}).
Based on preliminary experiments, we choose {\em ridge regression} as \diva's meta-learner.

Given a new and potentially poisoned input dataset, \diva uses the meta-learner to predict the clean accuracy and compares it to the empirical accuracy evaluated on the training data.
\diva identifies a dataset as poisoned if the
difference
between the empirical accuracy and the estimated accuracy exceeds a threshold (see Figure \ref{fig.diva}).
The accuracy difference itself quantifies the impact of the poisoning attack on the classifier.
%, i.e., a difference exceeding the threshold by a large margin indicates a strong effect.

The threshold for how many misleading labels are acceptable is a task-specific decision.
We suggest a heuristic for the threshold $t$ based on the maximum tolerated accuracy difference $\delta$ and the observed empirical accuracy $\text{acc}_\text{empirical}$,
$t = \text{acc}_{\text{empirical}}(\delta/100).$
%$$t = \text{acc}_{\text{empirical}}(\delta/100).$$
For example, if we set $\delta = 5\%$ and observe $90\%$ empirical accuracy, the threshold is $t=90\times(5/100)=4.5\%$.

\begin{figure}[t]
    \centering
    \includegraphics[width=0.99\columnwidth]{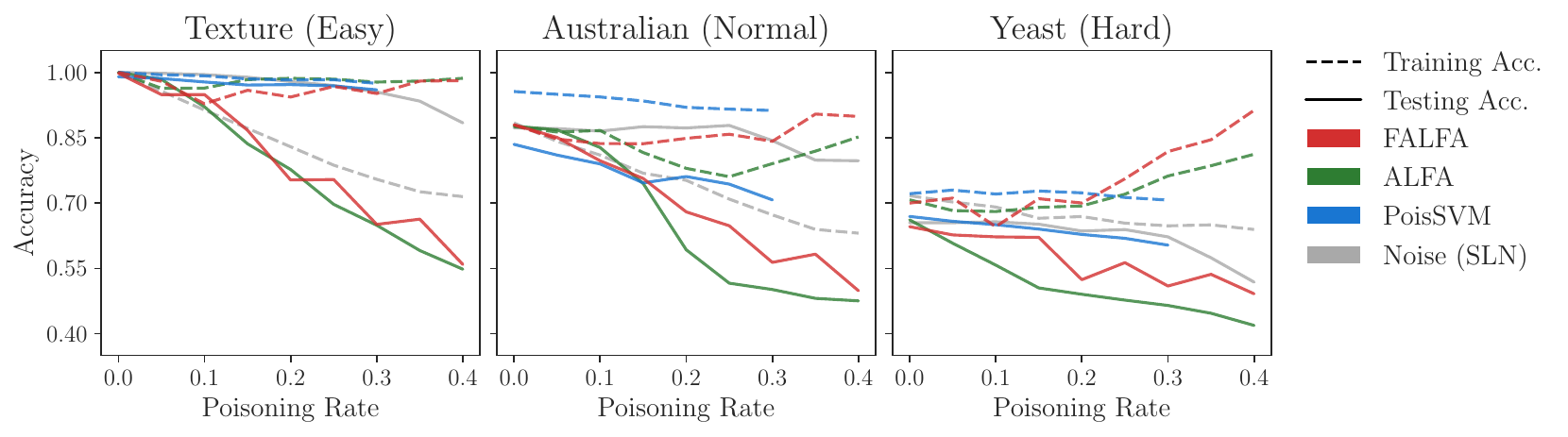}
    \caption{
    % Training and testing accuracies at various poisoning rates under different attacks.
    % The characteristics of the attack change based on the difficulty of the classification task.
        % \falfa, ALFA and SLN are used to attack neural network models, and PoisSVM is used to attack SVM models.
    The training and test accuracies at various poisoning rates exhibit similar patterns under different attacks on the same dataset. However, the difficulty of the classification task has a high influence on the behavior of attacks.
    }
    \label{fig.real_acc}
\end{figure}

\begin{figure}[t]
    \centering
    \includegraphics[width=0.99\columnwidth]{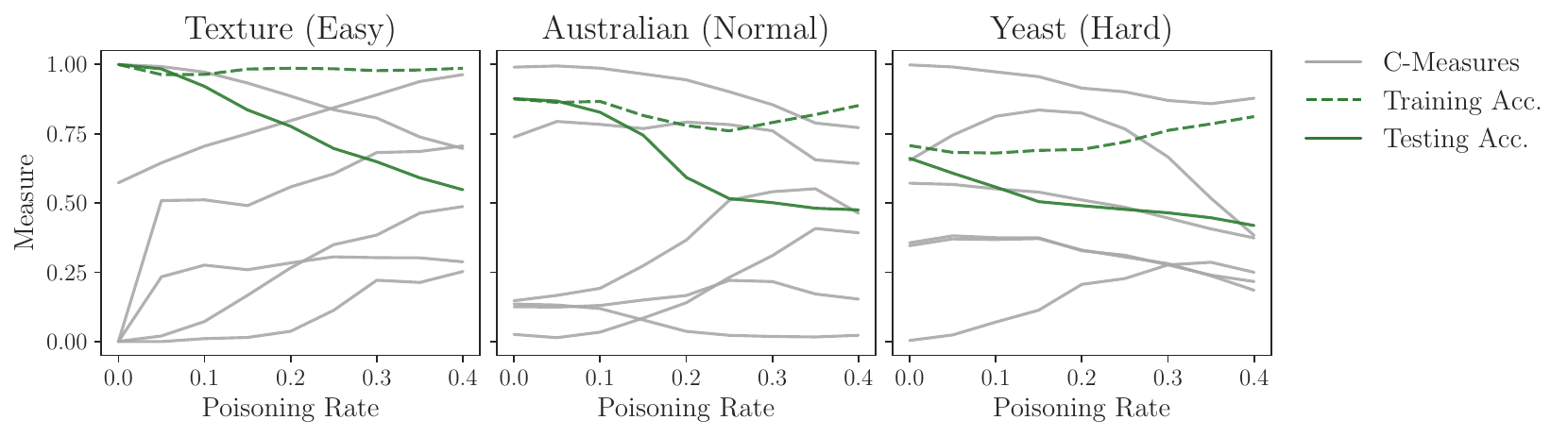}
    \caption{Under poisoning attacks (ALFA), C-Measures show positive
      and negative correlations to the clean testing accuracy
      (otherwise the lines would be flat), which is unknown at the
      training time, although the training accuracy holds almost
      flat. (Only selected measures are shown.)}  
    \label{fig:line}
\end{figure}

\tinytitle{Inference}
Once \diva fills a meta-database, trains a meta-learner, and selects a threshold, the training process is completed. 
To decide whether a new dataset is poisoned or not, \diva proceeds as in Figure~\ref{fig.diva}: 
First, it trains a classifier on the dataset and evaluates its empirical accuracy (typically using cross-validation for stable results). 
Second, \diva extracts C-Measures from the dataset to obtain a feature representation consistent with those in the meta-database. The pre-trained meta-learner can be applied to this representation to estimate the accuracy of the classifier on a clean dataset.
Third, \diva calculates the difference between both accuracy values and flags the dataset as poisoned if it exceeds the threshold and as clean otherwise.

Note that the meta-database is designed to be diverse such that it does not need to be constructed specifically for each dataset -- instead, in a real-world use-case, it will be collected only once and can continue to improve over time, making meta-learning a sustainable option despite the high up-front costs.

% The threshold for how many misleading labels are acceptable is a task-specific decision.
% We suggest a heuristic for the threshold $t$ based on the maximum tolerated accuracy difference $\delta$ and the observed training accuracy $\text{Acc}_\text{train}$
% $$t = \text{Acc}_{\text{train}}(\delta/100).$$
% For example, if we set $\delta = 5\%$ and observe $90\%$ training accuracy, then the threshold is $t=90\times(5/100)=4.5\%$.

% \subsection{Extending to Multi-Class Classification}
% So far, we present the algorithm in a binary classification setting.
% On a multi-class dataset, the attacker can target one class by applying the {\em one-versus-rest} (OVR) technique, merging other classes except the targeted one, resulting in the model misclassifying other classes as the targeted class.
% Similarly, we can use the OVR strategy when computing C-Measures on multi-class data. 
% A dataset is poisoned if any pairwise measure exceeds the threshold.

\section{Experiments}
\label{sec.exp}
% SUMMARY: Explain 2 parts of the experiments
We analyze each component used by \diva as well as
the entire framework. 
%, including different label poisoning attacks, C-Measures, and the overall performance of the framework.
Experiments were repeated 5 times to ensure robustness, and we report
arithmetic means. %unless otherwise specified.
To validate our proof-of-concept, we evaluate \diva on a large
meta-database of synthetic datasets and representative standard datasets and models. 
To ensure reproducibility, all data, hyperparameters, pre-trained
classifiers, results, code, and supplementary materials are available
at
\href{https://anonymous.4open.science/r/diva-aistats/}{\UrlFont{anonymous.4open.science/r/diva-data-poisoning}}.

\eventiniertitle{Datasets}
% SUMMARY: Real datasets
We evaluate \diva on 10 real-world datasets
%, i.e., Abalone, Australian, Banknote, Breastcancer, CMC, HTRU2, Phoneme, Ringnorm, Texture, and Yeast 
from the UCI ML repository \citep{Dua2019}. See our supplementary materials for details. 
% To reduce the running time, we downsample datasets to 2000 examples where applicable while keeping the {\em Positive Label Rate} (PLR) at $0.5$.
All datasets use an 80-20 training and testing split.
We use {\em leave-one-dataset-out} cross-validation to evaluate the
performance of \diva by training the meta-learner on 9 datasets and
evaluating on the remaining one. 
In practice, \diva needs to be
trained once on a representative set of datasets, attacks, and
models, and can be applied to any new dataset, new attack, or new
model without the need for re-training. 

%We also utilize the high level of control only synthetic data can offer when investigating specific properties of the framework.
% Additionally, we carry out tests on real datasets to demonstrate \diva's effectiveness in the wild.
%
We conducted two experiments using synthetic data.
The first one is based on difficulty, which we categorize
%In practice, the difficulty of a dataset depends on the specific classification task. 
%Here, we simplify this process by categorizing synthetic data 
as ``Hard'', ``Normal'' or ``Easy'' based on the testing accuracies of a nonlinear SVM being $<70\%$, $\geq70\%$ and $<90\%$, or $\geq90\%$, respectively.
Each difficulty level consists of 50 datasets.            
The second experiment is based on label noise.
We generate 50 datasets first and add uniformly distributed label noise up to $40\%$ in steps of $5\%$ resulting in 450 datasets for the noise-based experiment.
% All hyperparameters and generated datasets are available in our repository.

% SUMMARY: Classifier
\eventiniertitle{Attacks}
We include \falfa \citep{chang2024fast}, ALFA \citep{xiao2012adversarial} and Poisoning SVM (PoisSVM) \citep{biggio2012poisoning} as they are well-established representative attacks on binary classification. PoisSVM is largely different from \falfa allowing us to showcase \diva's generalizability to new attacks.
For \falfa, we train a neural network with 2 hidden layers (128 neurons each) using Stochastic Gradient Descent for a maximum of 400 epochs, learning rate $0.01$, and mini-batch size 128.
For ALFA and PoisSVM, we use SVM with an RBF kernel, and the parameters $C$ and $\gamma$ are tuned by a 5-fold CV.
We also include {\em Stochastic Label Noise} (SLN) as a baseline attack, which randomly flips a percentage of labels.
Moreover, PoisSVM is an {\em insertion attack}, so the poisoning rate is the percentage of additional examples added to the dataset.

\eventiniertitle{Parameters for \diva}
We select ridge regression as \diva's meta-learner and tune the
parameter $\alpha$  by a 5-fold CV. 
The threshold for \diva depends on the specific task.
Here, our goal is to evaluate \diva's effectiveness against various
attacks when a single threshold is selected without knowing the attack
algorithm. 
Hence, we select the threshold based on the {\em True Negative Rate}
(TNR) when testing on poison-free data. We provide results for
additional threshold settings in the supplementary material and our
repository.

\eventiniertitle{Baseline Detection}
We compare \diva with the kNN-based defense by \cite{paudice2018label} since it is an improvement on previous methods \citep{vorobeychik2018adversarial} and does not tie to a specific classification algorithm\footnote{We use the original implementation and parameters.}.
This defense is a point-wise sanitization technique. To tweak it towards dataset-wise decisions, once the algorithm relabels the training data, we compare the percentage of the difference between sanitized and original labels. 
%For the corresponding experiment (Section \ref{sec.attack.diva}), a single threshold is set based on $98\%$ TNR in consistency with \diva's setting there.
If the difference exceeds a threshold, the dataset is flagged as poisoned.
Note that we intentionally do not compare with poisoning detectors that require a clean sample as they do not operate under the same problem setting.

\tinytitle{Behavior of Attacks}
\label{sec.exp_attack}
% SUMMARY: Results on real data
% \diva is most effective when the attack has significant impact on the classifier's performance, whereas an attack that only leads to minor classification error might not trigger \diva, as explained in Section~\ref{sec.prob}.
\diva is based on the hypothesis that poisoning attacks can be
identified by comparing the poisoned and clean accuracies (Section \ref{sec.prob}), hence it is crucial to ensure the effectiveness of attacks.
Figure \ref{fig.real_acc} illustrates the relationship between the poisoned training and the clean test accuracy when a classifier is poisoned at various rates\footnote{We are only able to run PoisSVM up to 30\% poisoning rate using the implementation provided by the original paper \citep{biggio2012poisoning}.}.
We observe that \falfa and ALFA reliably degrade the classifier performance on all three datasets. However, PoisSVM is substantially weaker.

% Previous works use the test classification error as the sole performance metric to benchmark the poisoning attacks \citep{ho2002complexity,koh2022stronger,paudice2018label}.
% However, this does not fully reflect the impact of poisoned labels.
To conceal an attack from the user, keeping the training accuracy (dashed lines in Figure \ref{fig.real_acc}) stationery is equally essential.
Stochastic label noise (gray lines in Figure~\ref{fig.real_acc}) behaves differently from other attacks, as the training and test accuracies fall at a similar rate, highlighting the key motivation of this research.

% % SUMMARY: Talk about the speed
% When preparing the meta-database, \diva requires poison generators to inject poisoned examples at multiple levels.
% Thus, it is beneficial to choose efficient poisoning attacks.
% \falfa is substantially faster than other label poisoning attacks \citep{biggio2012poisoning,xiao2012adversarial,paudice2018label}.
% Overall, we noticed a speed-up of at least a factor of $20$ using \falfa.
% Full results are included in the supplementary material.

%\subsubsection{Edge Cases.}
Given a difficult classification task,
% as the poisoning rate increases, testing accuracy decreases until it closes to random guess, while training accuracy on poisoned data rises.
an increasing poisoning rate has a limited impact on the testing accuracy (Figure \ref{fig.real_acc} (right)).
Meanwhile, the training accuracy on poisoned data rises.
This pattern is observed in poisoning attacks, but not in noise.
% We believe this is due to the poisoning attacks optimizing in \falfa.
% When the testing accuracy is close to random levels, an attack algorithm can no longer maximize the loss on test data; it will minimize the loss on the training data instead.
% Thus, the classifier becomes more likely to overfit the poisoned training set.
\diva takes advantage of this characteristic.
%since the discrepancy between training and test accuracies still exists, despite the limited room for an increased classification error.
However, weak attacks may fail in this scenario, as seen in PoisSVM leading to only $6.6\%$ performance loss at a $30\%$ poisoning rate on Yeast. This case represents a Failed Attack (see Table \ref{tab:attack-matrix}), which \diva is not designed to detect.

\tinytitle{Evaluating C-Measures}
\label{sec.exp_cmeasures}
A classifier's performance depends on the quality of the features describing the examples.
Figure~\ref{fig:line} shows that C-Measures indeed carry information if a dataset is poisoned or not, and hence are a good choice as features for the meta-learner.
To improve readability, we only show the measures with the top-6 strongest fluctuations.
% When no poisoning attack is present, C-Measures strongly correlate to the classifier's testing accuracy.
% Most measures have a strong negative correlation to accuracy, except standard deviations, network measures, and dimensionality measures.
% This suggests that lower C-Measures indicate a dataset is less complex, and we should expect the classifier trained on this dataset to have a high performance.
%
When a dataset is attacked, we observe that C-Measures react to poisoned examples even at a low poisoning rate ($5\%$) when compared to the clean case (with poisoning rate $0\%$).
This indicates that C-Measures can identify poisoned data without access to the ground truth, even at a low poisoning rate.

% A common strategy of a white-box attack is to compute the gradient from a loss function which combines the losses of classifier and detector \citep{munoz2017towards}.
% Even if the attacker is aware \diva exists, computing the gradient of 28 loss functions is significantly more complex than a single loss function.

\tinytitle{Effectiveness of \diva}
\label{sec.attack.diva}
% This subsection investigates the entire framework's performance with regard to deciding whether a dataset is poisoned. We evaluate how well it is able to actually detect poisoning and compare it to a baseline method. 
%
% \eventiniertitle{Estimating Clean Testing Accuracy}
To demonstrate that \diva can estimate the clean accuracy on a poisoned dataset, we train the meta-learner on datasets poisoned by SLN and \falfa, and test on unseen datasets with known (\falfa) and unknown attacks (ALFA and PoisSVM).
Table \ref{table.real_rmse} provides the {\em Root Mean Square Error} (RMSE) of this regression.
The RMSEs on unknown attacks are comparable to known attacks, suggesting that \diva's meta-learner can make accurate predictions even on unseen datasets attacked by unknown attacks.

\begin{table}[t]
    \scriptsize
    \newcommand\mypm{\mathord{\pm}}
    \centering
    \caption{Root Mean Square Error (RMSE) of \diva estimating clean testing accuracy when trained on \falfa and SLN.
        \diva performs consistently even on unseen attacks.
        (Difficulties are based on the testing accuracy of clean data -- \underline{E}asy/\underline{N}ormal/\underline{H}ard.)
    }
    \setlength{\tabcolsep}{0.5em}         % adjust horizontal padding here
    \begin{tabular}
    %{l|cccc}
    {l@{\hspace{0.3em}}c@{\hspace{0.5em}}c@{\hspace{0.5em}}c@{\hspace{0.5em}}c}
        \toprule
        Dataset       & SLN                    & \falfa                 & PoisSVM                & ALFA                   \\
        &  (noise) & (known) & (unknown) & (unknown)\\
        \midrule
        Banknote (E)            & $0.16 \mypm 0.04$        & $0.15 \mypm 0.05$        & $0.17 \mypm 0.03$        & $0.09 \mypm 0.04$        \\
        Breastcancer (E)         & $0.07 \mypm 0.03$        & $0.11 \mypm 0.01$        & $0.07 \mypm 0.02$        & $0.12 \mypm 0.04$        \\
        HTRU2 (E)                & $0.06 \mypm 0.01$        & $0.09 \mypm 0.01$        & $0.06 \mypm 0.01$        & $0.11 \mypm 0.03$        \\
        Ringnorm (E)            & $0.19 \mypm 0.07$        & $0.13 \mypm 0.02$        & $0.15 \mypm 0.05$        & $0.14 \mypm 0.04$        \\
        Texture (E)              & $0.09 \mypm 0.03$        & $0.07 \mypm 0.02$        & $0.10 \mypm 0.04$        & $0.07 \mypm 0.03$        \\
        Abalone (N)              & $0.07 \mypm 0.02$        & $0.08 \mypm 0.01$        & $0.06 \mypm 0.02$        & $0.11 \mypm 0.06$        \\
        Australian (N)          & $0.07 \mypm 0.03$        & $0.11 \mypm 0.03$        & $0.06 \mypm 0.02$        & $0.11 \mypm 0.04$        \\
        CMC (N)                 & $0.10 \mypm 0.02$        & $0.21 \mypm 0.06$        & $0.25 \mypm 0.08$        & $0.25 \mypm 0.11$        \\
        Phoneme (N)             & $0.09 \mypm 0.03$        & $0.09 \mypm 0.03$        & $0.07 \mypm 0.02$        & $0.08 \mypm 0.02$        \\
        Yeast (H)              & $0.08 \mypm 0.04$        & $0.12 \mypm 0.02$        & $0.09 \mypm 0.04$        & $0.19 \mypm 0.05$        \\
        \midrule
        \textbf{Mean}      & \textbf{0.10$\mypm$0.04} & \textbf{0.12$\mypm$0.04} & \textbf{0.11$\mypm$0.06} & \textbf{0.12$\mypm$0.06} \\
        \bottomrule
    \end{tabular}
    \label{table.real_rmse}
\end{table}

\begin{figure}[t!]
    \centering
    \includegraphics[width=0.825\columnwidth]{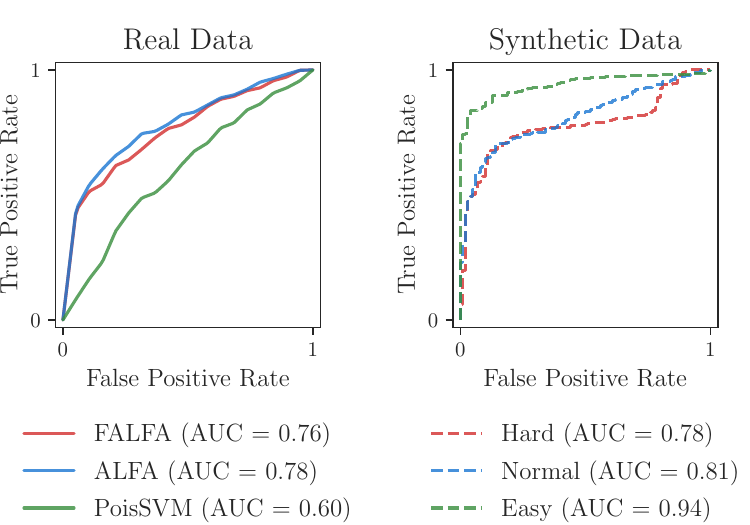}
    \caption{ROC curve for \diva's prediction on whether the training set is poisoned.
        Left: Unseen real datasets poisoned by FALFA (known to \diva), ALFA, and PoisSVM (unknown).
        %(left): Unseen real datasets poisoned by FALFA, ALFA, and PoisSVM. ALFA and PoisSVM are unknown attacks to \diva.
        Right: Synthetic datasets poisoned by FALFA; grouped by dataset difficulty.}
    \label{fig.roc}
\end{figure}

\eventiniertitle{Performance on Real Data}
We generated {\em Receiver Operating Characteristic}  (ROC) curves to examine how well \diva can distinguish whether a dataset is poisoned as its threshold is varied (Figure \ref{fig.roc} (left)).
We estimate the empirical-clean accuracy differences and then vary the threshold to compute the curves.
%We compute the curves by estimating the empirical-clean accuracy differences first and then varying the threshold.
To show that \diva is effective against unseen attacks, we poison all real-world datasets with the known attack \falfa and the unknown ALFA and PoisSVM.

\begin{figure}[t!]
    \centering
    \includegraphics[width=0.99\columnwidth]{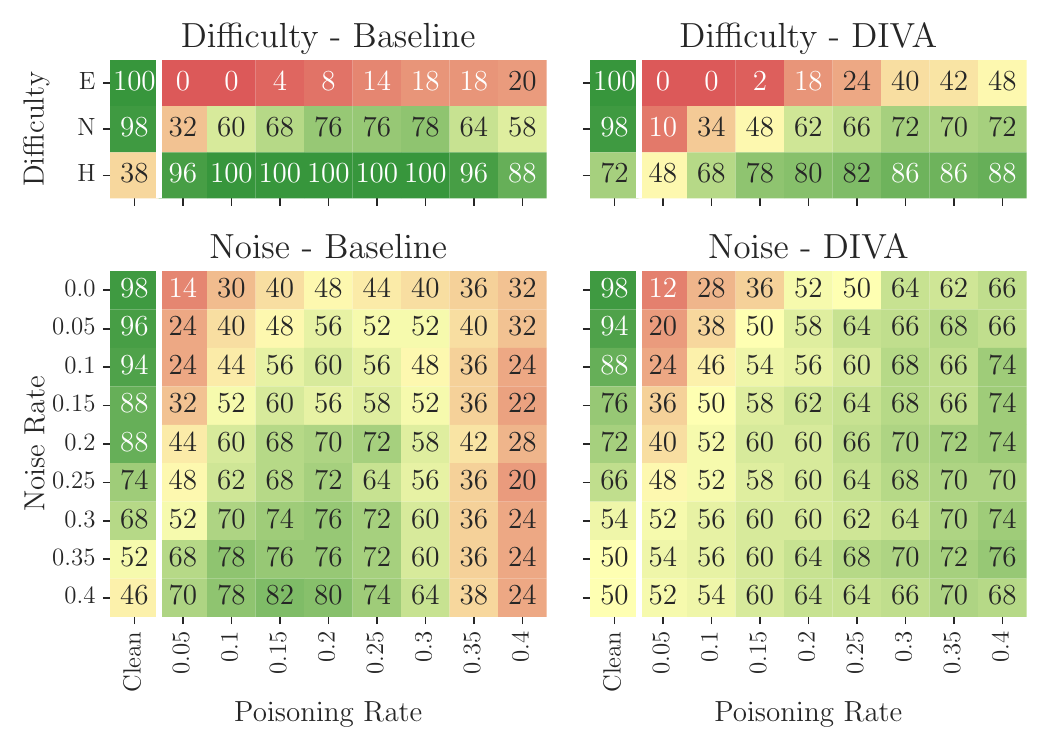}
    \caption{Performance of the baseline (kNN-based defense \citep{paudice2018label}) and \diva on synthetic data when a single threshold is set at $98\%$ TNR. The columns labeled ``Clean'' show TNRs (ratio of clean datasets identified as clean) as there are no poisoning labels. The remaining columns show TPRs (poisoned datasets identified as poisoned).}
    \label{fig:heatmap}
\end{figure}

%The ROC AUCs for \falfa, ALFA and PoisSVM are 0.76, 0.78 and 0.60, respectively.
The ROC AUCs on \falfa and ALFA are similar, which suggests that \diva performs consistently on both known and unknown attacks.
PoisSVM has a lower ROC AUC score, although Table \ref{table.real_rmse} indicates that \diva can successfully predict clean accuracy when a dataset is poisoned with PoisSVM.
%PoisSVM has a lower ROC AUC score than the others, although the RMSE on PoisSVM indicates that \diva can successfully predict the clean accuracy when a dataset is poisoned (Table \ref{table.real_rmse}).
We attribute the lower ROC AUC score on PoisSVM to the following:
\begin{enumerate*}
    \item PoisSVM fails to degrade the classifiers' performance on half the datasets we tested.
    \item The meta-learner has only trained on NN models but PoisSVM attacks on SVM models, which have lower baseline accuracy (Figure \ref{fig.real_acc}).
\end{enumerate*}
\diva prioritizes detecting poisoning attacks that significantly impact the classifier performance rather than less effective ones that have minimal impact (Section \ref{sec.prob}).
A larger and more diverse meta-database can mitigate this issue but was out-of-scope for this proof-of-concept paper.

\eventiniertitle{Experiments on Synthetic Data Based on Difficulty}
To demonstrate that \diva is effective regardless of the difficulty, we test \diva on our synthetic data grouped by difficulties (see  Figure \ref{fig.roc} (right)). 
%The ROC AUCs for easy, normal and hard datasets are 0.94, 0.81 and 0.78, respectively.
% The ROC AUCs on \falfa and ALFA are similar on synthetic data when testing on normal and hard data, further underlining that the lower bound of \diva's performance exists regardless of the difficulty.
%
\diva is most effective when the classifier has high performance, as the resulting accuracy drop is larger.
% In comparison, \diva's performance on real-world datasets is similar to synthetic data with hard difficulty, indicating that \diva works even when the classification task is difficult where the class imbalance and overlapping features exist.

To further analyze \diva's performance,
we compare it with the baseline.
The ``Clean" columns of Figure \ref{fig:heatmap} contain only clean training sets and display TNRs as there are no poisoning labels.
%The first columns of Figure \ref{fig:heatmap}, labeled with ``Clean'', contain only the clean training sets and display {\em True Negative Rates} (TNRs) (as there are no poisoning labels).
The other cells contain datasets with various percentages of poisoning examples and hence display {\em True Positive Rates} (TPRs).

We make three observations in Figure \ref{fig:heatmap} (top):
\begin{enumerate*}
    \item \diva is monotonic with respect to the poisoning rate, while the performance of the baseline falls at higher rates. Stronger attacks are easier to detect for \diva.
    \item The choice of the threshold is important to accommodate different dataset difficulties. %The threshold was selected for a normal difficulty here. 
    We selected the threshold for a normal difficulty; a lower and higher value would be more suited for hard and easy datasets, respectively.
    %A lower value would be more suited for the hard difficulty, and a higher threshold is required for easy datasets.
    \item The baseline has almost half \diva's TNR on hard datasets, suggesting that it cannot distinguish between hard and poisoned data whereas \diva performs monotonically across all difficulties.
\end{enumerate*}

\eventiniertitle{Distinguishing Between Noisy and Poisoned Data}
To demonstrate that \diva can differentiate between noisy and poisoned data, we selected a single threshold at $98\%$ TNR on the noise-free dataset and used it throughout the experiment. 
The results for the baseline and \diva are presented in Figure \ref{fig:heatmap} (bottom). 
In the first column, we observe an increase in FPRs as label noise increases, which is expected since noisy datasets are more challenging to classify and the threshold is not adjusted to compensate for the increasing noise rates.
The first row in both heatmaps represents the detectors' performance when there is no noise. 
\diva shows similar performance compared with the ``Normal'' difficulty in Figure \ref{fig:heatmap} (top). 
However, the baseline performs substantially worse than \diva, due to mixed difficulties in this experiment and the baseline's specialization to hard datasets (as seen in Figure \ref{fig:heatmap} (top)).
This further demonstrates that \diva performs consistently regardless of the data's difficulties.

Similar to the heatmap for difficulties, \diva behaves monotonically with respect to both the poisoning and the noise rate. 
The highest TPRs are reported at the highest poisoning rates, except for the largest noise rate. 
Meanwhile, the baseline's performance starts to fall off after $25\%$ noise. The highest TPR for the baseline is reported at $40\%$ noise with $15\%$ poisoned data. This suggests that \diva's capability is not generally compromised by noisy data, whereas the baseline cannot distinguish between noisy and poisoned data.
Overall, \diva performs consistently well, even in the worst scenario where noisy data contains a high percentage of poisoned examples.

\section{Related Work}
\label{sec.relatedwork}

Besides \diva, no poisoning attack detector has been proposed yet in the area of automated detection of data quality issues without a clean sample \citep{Dost2020,Bellamy2019,chang2021baard}.
%Automated detection of data quality issues without a clean sample has gained recent attention \citep{Dost2020,Bellamy2019,chang2021baard}, yet
This section reviews related work to data poisoning attacks, defenses, and existing applications for C-Measures.

% SUMMARY: Attackers' goals are all the same regardless the attacks.

\tinytitle{Attacks}
% The studies on data poisoning attacks are divided in two directions based on the adversary's goals: {\em error-generic} and {\em error-specific}. 
% Error-generic poisoning attacks were the earliest attacks targeting PCA \citep{rubinstein2009antidote} and SVM \citep{biggio2012poisoning}, but later studies revealed that logistic regression, linear regression, and DNN were also vulnerable \citep{mei2015using,munoz2017towards}. 
% Backdoor attacks, where the adversary plants a trigger into the classifier, have been proposed recently \citep{gu2017badnets}. 
% While recent studies focus on backdoor attacks with clean labels \citep{li2021anti,shokri2020bypassing,liu2018fine,hayase2020spectre,wang2021robust}, only a few works explore label poisoning attacks despite their high accessibility \citep{zhang2020adversarial,rosenfeld2020certified}.
%
Data poisoning attacks can be categorized based on the attacker's goal, either compromising availability or integrity \citep{cina2022wild}.
Among availability attacks, label flipping attacks are the most accessible to attackers, but challenging to optimize. A recent survey on poisoning \citep{cina2022wild} lists four label flipping attacks: 
\citet{biggio2012poisoning} proposed an algorithm to craft adversarial labels for attacking SVMs.
\citet{zhang2020adversarial} and \citet{paudice2018label} extended label-flipping attacks to graph neural networks and neural networks, respectively.
\citet{xiao2012adversarial} proposed ALFA and later \citet{xiao2015feature} improved the algorithm. ALFA finds the label flip that maximizes the classification error and is included in our experiments. 
While recent studies focus on backdoor attacks with clean labels \citep{li2021anti,shokri2020bypassing,liu2018fine,hayase2020spectre,wang2021robust, gu2017badnets}, only a few works explore label poisoning attacks despite their high accessibility \citep{zhang2020adversarial,rosenfeld2020certified, cina2022wild}.

% While recent research has predominantly focused on integrity attacks, such as backdoor attacks with clean labels \citep{gu2017badnets, shokri2020bypassing, li2021anti}, there has been limited exploration of the possibility of label flipping attacks on neural network models \citep{rosenfeld2020certified, cina2022wild}.

% SUMMARY: Limitation on point-wise sanitization
\tinytitle{Defenses}
Many data poison defenses are based on point-wise sanitization, such
as subsampling on PCA \citep{rubinstein2009antidote}, random smoothing
on partitions of the training set \citep{levine2020deep}, anomaly
detection \citep{paudice2018detection}, and label
reassignment based on kNN \citep{paudice2018label}.  
\citet{koh2022stronger} showed that a strong poisoning
attack can break data sanitization. 
Another approach is robust learning. These methods modify the
optimization process \citep{wang2021robust,li2021anti} or apply neuron
pruning on DNNs \citep{wang2019neural,liu2018fine}. 
However, all of these methods are weak against strong poisoning
attacks and specific to certain classifiers
\citep{koh2022stronger,wang2021robust,wang2019neural,levine2020deep,chen2021pois,chen2018detecting,wang2020practical}.
Other attempts to detect attacks are based on performance metrics or
comparisons between a trusted subset and the training set
\citep{mozaffari2014systematic,baracaldo2017mitigating,wu2021indirect}.  
A recent survey \citep{yang2023application} highlights the promising
results of meta-learning models in cybersecurity, particularly in
intrusion and anomaly detection. However, existing meta-learning
approaches for data poisoning detection are typically limited to NNs
\citep{xu2021detecting,kolouri2020universal}. 
FLTrust \citep{cao2022fltrust} is a recent federated learning method
that mitigates the impact of malicious clients by assigning trust
scores to incoming datasets. 
In contrast, \diva neither requires a clean subset, nor depends on
a specific type of classifier, and is robust against high poisoning
rates.

Before sanitizing data, it is important to recognize potential
threats. Existing paradigms like certified defenses provide insights
into a model's performance under hypothetical attacks but do not offer
a straightforward solution to detect poisoned data
\citep{steinhardt2017certified,rosenfeld2020certified}. \diva is a
framework that provides a reliable and direct answer to detecting label-poisoning attacks.
%, and it can be applied before and after sanitizing
%data to ensure that the dataset is free from poisoning attacks. 

% SUMMARY: Related work on C-Measures
\tinytitle{C-Measures}
C-Measures have shown to be valuable in numerous fields, e.g., in
noise filter selection \citep{saez2013predicting}, instance selection
for reducing training time and removing outliers \citep{leyva2014set},
and dynamic classifier selection for benchmarking classification
models \citep{brun2018framework}.  
However, we are the first to link them to poisoning attacks.

\section{Conclusions and Future Work}
\label{sec.conclusion}
Poisoning attacks corrupt a classifier's training data, causing a drop in performance.
Existing defenses typically require clean data, which is often unavailable, or are not fully-agnostic.
We propose \diva, a novel fully-agnostic meta-learning framework
that detects poisoning attacks by estimating the dataset's clean
accuracy using meta-data, overcoming the need for clean data. \diva is
the first fully-agnostic algorithm that, on a dataset level, can reliably predict if
it is poisoned or not, without the need for a ground truth, only by
examining the potentially poisoned data. This in itself is a big
advancement in the field of data poisoning, which relies on our
observation that we can exploit a relationship between C-Measures and
data poisoning.   
Our experiments confirm \diva's ability to detect poisoning threats in
unseen datasets, making it valuable for automated data quality
surveillance while also making machine learning models less vulnerable
to poisoning attacks. Future work will explore the
diversification of the meta-database in terms of attack and data types
to assess if one generalizable meta-learner is superior to an ensemble of
specialized meta-learners or vice-versa. 
We will make our meta-database publicly available to aid future
research.
% expand the framework to other poisoning attacks, e.g., backdoor attacks.
% Future work will expand the framework to other data types, such as images and time-series data, to further enhance its applicability.

%%%%%%%%%%%%%%%%%%%%%%%%%%%%%%%%%%%%%%%%%%%%%%%%%%%%%%%%%%%%%%%%%%%%%%%%%%%%%%%

\bibliographystyle{apalike}
\bibliography{references.bib}

\clearpage


\section*{Supplementary material}

# Supplementary Material
# Poison is Not Traceless: Fully-Agnostic Detection of Poisoning Attacks

## Reproducibility Checklist

1. For all models and algorithms presented, check if you include:
   (a) A clear description of the mathematical setting, assumptions, algorithm, and/or model. [Yes]
   (b) An analysis of the properties and complexity (time, space, sample size) of any algorithm. [Yes, in the supplementary material]
   (c) (Optional) Anonymized source code, with specification of all dependencies, including external libraries. [Yes]

2. For any theoretical claim, check if you include:
   (a) Statements of the full set of assumptions of all theoretical results. [Not Applicable]
   (b) Complete proofs of all theoretical results. [Not Applicable]
   (c) Clear explanations of any assumptions. [Not Applicable]

3. For all figures and tables that present empirical results, check if you include:
   (a) The code, data, and instructions needed to reproduce the main experimental results (either in the supplemental material or as a URL). [Yes, in the anonymous repository]
   (b) All the training details (e.g., data splits, hyperparameters, how they were chosen). [Yes, in the supplementary material]
   (c) A clear definition of the specific measure or statistics and error bars (e.g., with respect to the random seed after running experiments multiple times). [Yes, seed values are provided in the anonymous repository]
   (d) A description of the computing infrastructure used. (e.g., type of GPUs, internal cluster, or cloud provider). [Yes, in the supplementary material]

4. If you are using existing assets (e.g., code, data, models) or curating/releasing new assets, check if you include:
   (a) Citations of the creator If your work uses existing assets. [Yes]
   (b) The license information of the assets, if applicable. [Not Applicable, packages are based on BSD-style license]
   (c) New assets either in the supplemental material or as a URL, if applicable. [Yes]
   (d) Information about consent from data providers/curators. [Not Applicable, packages are based on BSD-style license]
   (e) Discussion of sensible content if applicable, e.g., personally identifiable information or offensive content. [Not Applicable]

5. If you used crowdsourcing or conducted research with human subjects, check if you include:
   (a) The full text of instructions given to participants and screenshots. [Not Applicable]
   (b) Descriptions of potential participant risks, with links to Institutional Review Board (IRB) approvals if applicable. [Not Applicable]
   (c) The estimated hourly wage paid to participants and the total amount spent on participant compensation. [Not Applicable]

## Introduction

To ensure reproducibility, we provide additional details for the paper in this section. We include details on FALFA (including its time complexity), the hardware and software configurations for our experiments, detailed dataset descriptions, and additional results.

## Details of Falfa

In a poisoning attack, the attacker can directly manipulate the training data $\mathcal{D}_{\text{tr}} := \{(\mathcal{X}_{\text{tr}}, \mathcal{Y}_{\text{tr}})\}$ and interact with the training process of a model $f$. Based on this information, the attacker creates a poisoned training set $\mathcal{D}'_{\text{tr}} := \{(\mathcal{X}'_{\text{tr}}, \mathcal{Y}'_{\text{tr}})\}$, on which a poisoned model $f'$ is trained, by modifying or injecting examples into $\mathcal{D}_{\text{tr}}$. To increase the test error, the attacker's goal is to maximize the loss on the clean test set $\mathcal{D}_{\text{te}} := \{(\mathcal{X}_{\text{te}}, \mathcal{Y}_{\text{te}})\}$.

The attack objective of a label-flipping attack is:

$$\min_{\mathcal{D}'_{\text{tr}}} \ell(\mathcal{D}'_{\text{tr}}, f') - \ell(\mathcal{D}_{\text{te}}, f'),$$

which can be interpreted as the adversary making the discrepancy between $f$ and $f'$ as large as possible.

FALFA is derived from Equation , which we rewrite with the adversary's cost constants as:

$$\begin{aligned} \min_{\mathcal{Y}'_{\text{tr}}} \quad & \ell(f'(\mathcal{X}_{\text{tr}}), \mathcal{Y}'_{\text{tr}}) - \ell(f(\mathcal{X}_{\text{tr}}), \mathcal{Y}'_{\text{tr}}) \\ \text{s.t.} \quad & \sum_{i=1}^{n} |y'_i - y_i| \leq n\epsilon, \\ & y'_i \in \{0, 1\} \text{ for } i = 1, \ldots, n. \end{aligned} \quad (1)$$

A *Neural Network* (NN) optimizes the cross entropy loss $\ell(p_i, y_i) = y_i[-\log(p_i) + \log(1 - p_i)]$ and normalizes its outputs by the softmax function $p_i = \exp(x_i)/\sum_j \exp(x_j)$. In this case, we can transform Equation 1 into a linear programming problem in three steps:

1. We expand the objective function in Equation 1 using the cross entropy loss to $\ell(f'(\mathcal{X}_{\text{tr}}), \mathcal{Y}'_{\text{tr}}) - \ell(f(\mathcal{X}_{\text{tr}}), \mathcal{Y}'_{\text{tr}}) = (\alpha - \beta)\mathcal{Y}'_{\text{tr}}$ with $\alpha := -\log(p') + \log(1 - p')$ and $\beta := -\log(p) + \log(1 - p)$. $\beta$ only depends on the original classifier $f$ and clean data $\mathcal{D}_{\text{tr}}$, so it is a constant and can be computed beforehand.

2. Since the problem is a binary classification, we simplify the condition $\sum_{i=1}^{n} |y'_i - y_i| \leq n\epsilon$ to $\lambda \cdot \mathcal{Y}'_{\text{tr}} \leq n\epsilon + \lambda \cdot \mathcal{Y}_{\text{tr}}$ using a multiplier $\lambda = 1$ if $y_i = 0$ and $-1$ otherwise. $\lambda$ is a constant vector as it only depends on $\mathcal{Y}_{\text{tr}}$. We obtain the multiplier $\lambda$ by generalizing all the combinations between $y_i$ and $y'_i$ in a binary classification task, as shown in Table 1.

3. Using the simplex method, we relax the boundary condition $y'_i \in \{0, 1\}$ to $0 \leq y'_i \leq 1$, because the solution is guaranteed on the edges.

Overall, Equation 1 becomes the following linear programming problem:

$$\begin{aligned} \min_{\mathcal{Y}'_{\text{tr}}} \quad & (\alpha - \beta) \mathcal{Y}'_{\text{tr}} \\ \text{s.t.} \quad & \lambda \cdot \mathcal{Y}'_{\text{tr}} \leq n\epsilon + \lambda \cdot \mathcal{Y}_{\text{tr}}, \\ & 0 \leq y'_i \leq 1 \text{ for } i = 1, \ldots, n. \end{aligned} \quad (2)$$

The full algorithm of FALFA is shown in Algorithm 1.

FALFA is a suitable poison generator for DIVA in the following ways:

1. It is efficient on models using cross-entropy loss. Apart from the poisoned classifier $f'$ itself, only $p'$ and $\alpha$ require updates. Anything else is constant and can be computed beforehand.

Table 1: All combinations for $|y'_i - y_i|$. By introducing a multiplier $\lambda$, $\lambda \cdot (y'_i - y_i)$ is equivalent to $|y'_i - y_i|$.

| $y_i$ | $y'_i$ | $|y'_i - y_i|$ | $\lambda$ |
|---|---|---|---|
| 0 | 0 | 0 | 1 |
| 0 | 1 | 1 | 1 |
| 1 | 0 | -1 | -1 |
| 1 | 1 | 0 | -1 |

**Algorithm 1** FALFA (Fast Adversarial Label Flipping Attack)

**Require:** Original training set $\mathcal{D}_{tr} := \{(\mathcal{X}_{tr}, \mathcal{Y}_{tr})\}$, budget parameter $\epsilon$, classifier $f$ trained on $\mathcal{D}_{tr}$
**Ensure:** Poisoned training labels $\mathcal{Y}'_{tr}$
1: $p \leftarrow \text{Softmax}(f(\mathcal{X}_{tr}))$;
2: $\beta \leftarrow -\log(p) + \log(1-p)$;
3: $\lambda \leftarrow 1$ if $(y_i == 1)$ else $-1, \forall y_i \in \mathcal{Y}_{tr}$;
4: $\mathcal{Y}'_{tr} \leftarrow$ Randomly flip $n \cdot \epsilon$ examples on $\mathcal{Y}_{tr}$;
5: $f' \leftarrow f$;
6: **while** $\mathcal{Y}'_{tr}$ does not converge **do**
7:     Retrain classifier $f'$ using $(\mathcal{X}_{tr}, \mathcal{Y}'_{tr})$;
8:     $p' \leftarrow \text{Softmax}(f'(\mathcal{X}_{tr}))$;
9:     $\alpha \leftarrow -\log(p') + \log(1-p')$;
10:    Update $\mathcal{Y}'_{tr}$ by solving Eq. 2;
11: **return** $\mathcal{Y}'_{tr}$;

2. The computation time is invariant to poisoning rates. Because Equation 2 is solved by a linear programming solver, we update $\mathcal{Y}'_{tr}$ without looping through all permutations.

3. It is guaranteed solvable at any poisoning rate since we set the initial solution for $\mathcal{Y}'_{tr}$ by randomly flipping $n \cdot \epsilon$ examples.

4. FALFA is fast (we noticed a speed-up of at least factor 20 in our experiments; see our supplementary) which facilitates using a large meta-database.

## Time Complexity of Falfa

FALFA is more computationally efficient than ALFA and PoisSVM by a substantial margin. Linear programming is an exponential-time algorithm, the time complexity is around $O(n^{2.5})$. Xiao *et al.*'s ALFA creates a copy of $\mathcal{Y}_{tr}$ in the linear programming step, so $n$ is essentially doubled. Paudice *et al.*'s ALFA on NN is slower than Xiao *et al.*'s since it traverses all combinations of $\mathcal{Y}_{tr}$ instead of using linear programming. FALFA uses linear programming but without doubling $\mathcal{Y}_{tr}$, resulting in an approximately $2^{2.5} \approx 5.6$ times faster than ALFA on each iteration. Our test shows that FALFA converges at two iterations on average, but ALFA takes more than five iterations to converge. In the worst-case scenario, FALFA poisons the CMC dataset in $22.4 \pm 8.6$ secs, while ALFA requires $405.8 \pm 348.4$ secs, and PoisSVM took over 2 hours. We observe the minimal difference on Breastcancer, where FALFA completes the task at $5.3 \pm 1.9$ secs, and it takes ALFA $7.4 \pm 5.6$ secs.

## Hardware and Software Configurations

All experiments are conducted on a workstation with the following configurations:

- CPU: AMD Ryzen 9 5900 24 threads @ 4.4GHz
- GPU: Nvidia GeForce RTX 3090 24GB
- Memory: 64GB
- Operating System: Ubuntu 20.04.3 LTS

- Software: Python 3.8.10, PyTorch 1.10.1+cu113, scikit-learn 1.0.2

The baseline data poisoning attacks are obtained from *Adversarial Robustness Toolbox* (ART) 1.9.1 [1] and *Secure and Explainable Machine Learning in Python* (SecML) 0.15 [2].

The execution time mentioned in the paper is evaluated using the environment above.

**Real-World Datasets.**

All datasets are obtained from the UCI Machine Learning Repository [3]. We apply standardization on all datasets during the preprocessing.

For multi-class classification tasks, we convert the dataset into binary based on the following:

- **Abalone:** If the 'Rings' attribute is less than 10, we assign the example to the negative class; Else, assign to the positive class. We exclude the categorical attribute – 'Sex' and the output label – 'Rings' from the inputs.
- **CMC:** has 3 output classes: 1. No-use, 2. Long-term, and 3. Short-term. If the class is 'No-use', assign it to the negative class; Else, to the positive class.
- **Texture:** It has 10 output classes. We use a subset which contains examples labeled as '3' and '9'. If the class is '3', assign it to the negative class; Else, to the positive class.
- **Yeast:** It has 10 output classes. We select ''0 and '7', the top two classes sorted by sample size. If the class is '0', assign it to the negative class; Else, to the positive class.

**Synthetic Datasets.**

Table 2 shows the parameters we used to generate synthetic datasets.

Table 2: Hyper-parameters for generating synthetic data

| Parameter | Range |
| --- | --- |
| Sample size | $\{1000, 1001, \ldots, 2000\}$ |
| # of informative features | $\{4, 5, \ldots, 30\}$ |
| # of redundant features | $\{0, 1, \ldots, 5\}$ |
| Class ratio | $[0.4, 0.6]$ |

## Additional Results

**Classifiers' Performance.**

Table 3 shows the performance of classifiers trained on poison-free data.

**Performance Loss.**

Fig. 1 shows the performance loss on all real datasets.

**C-Measures on Clean and Poisoned Data.**

When no poisoning attack is present, the C-Measures strongly correlate to the classifier's test accuracy as can be seen in Fig. 2a. When the dataset has been poisoned, the C-Measures react to it. Despite a performance drop on the training accuracy of $1.0 \pm 5.6\%$, we observe a correlation drop across all measures when measuring on poisoned data, as shown in Fig. 2b. This matches the test accuracy drop, which indicates the data becomes more complex, despite only minor changes in the training accuracy.

---

[1]Source: https://github.com/Trusted-AI/adversarial-robustness-toolbox
[2]Source: https://github.com/pralab/secml
[3]Source: https://archive.ics.uci.edu/ml

Table 3: Summary of the training set size ($n$), number of features ($m$), Positive Label Rate (PLR), average accuracy (%) for training and test sets across all classifiers, and difficulty (Easy/Normal/Hard).

| Dataset | $n$ | $m$ | PLR | Train Acc. | Test Acc. | Diff. |
|---|---|---|---|---|---|---|
| Abalone | 1600 | 7 | 0.50 | $79.9 \pm 0.7$ | $76.5 \pm 0.5$ | N |
| Australian | 552 | 14 | 0.45 | $91.5 \pm 3.1$ | $81.9 \pm 2.1$ | N |
| Banknote | 1097 | 4 | 0.44 | $100.0 \pm 0.0$ | $100.0 \pm 0.0$ | E |
| Breastcancer | 455 | 30 | 0.63 | $99.3 \pm 0.2$ | $95.0 \pm 2.5$ | E |
| CMC | 1178 | 9 | 0.77 | $79.9 \pm 2.8$ | $77.5 \pm 0.6$ | N |
| HTRU2 | 1600 | 8 | 0.50 | $94.8 \pm 0.5$ | $92.6 \pm 0.9$ | E |
| Phoneme | 1600 | 5 | 0.50 | $89.7 \pm 6.3$ | $85.6 \pm 1.3$ | N |
| Ringnorm | 1600 | 20 | 0.50 | $99.4 \pm 0.4$ | $97.8 \pm 1.1$ | E |
| Texture | 800 | 40 | 0.50 | $100.0 \pm 0.0$ | $99.8 \pm 0.5$ | E |
| Yeast | 713 | 8 | 0.48 | $73.5 \pm 4.7$ | $65.8 \pm 1.6$ | H |

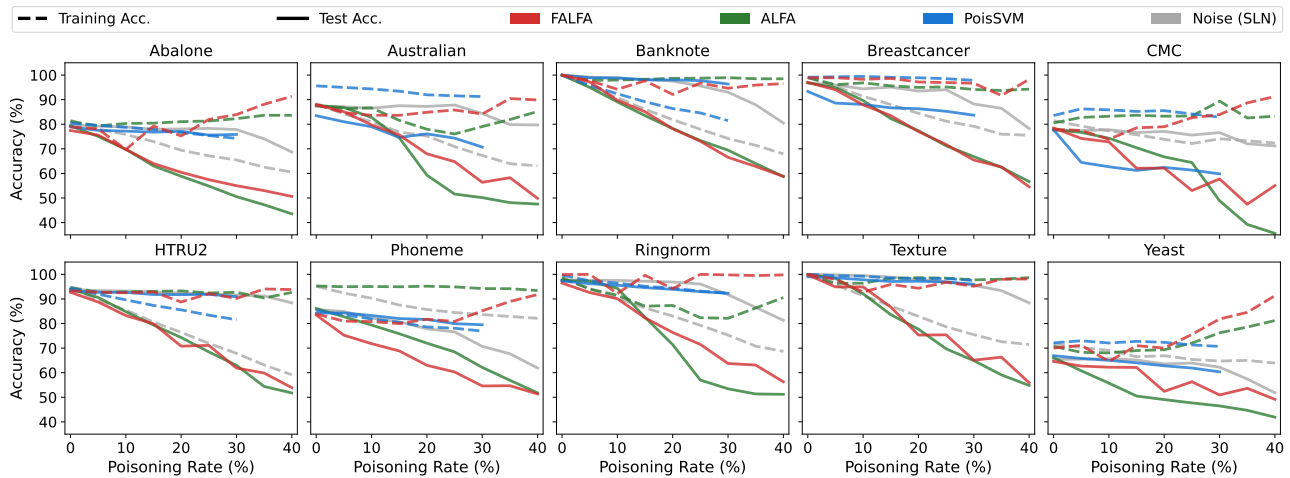

Figure 1: Train and test accuracy at various poisoning rates when classifiers under SLN, PoisSVM, and FALFA attacks.

**Performance Loss at a Low Poisoning Rate.**

Here, we present the performance loss at a low poisoning rate (10%) in Table 4. This is the test accuracy difference before and after the attack. PoisSVM has no meaningful impact ($< 2\%$) on the classifiers' performance in 7 out of 10 datasets. Meanwhile, SLN leads to minor performance improvement on CMC and Yeast.

Table 4: Performance loss (%) after attacked by a poisoning attack with 10% poisoning rate.

| Dataset | SLN | PoisSVM | ALFA | FALFA |
|---|---|---|---|---|
| Abalone | $0.8 \pm 0.7$ | $1.8 \pm 0.8$ | $9.5 \pm 1.9$ | $7.7 \pm 1.7$ |
| Australian | $0.7 \pm 0.5$ | $4.5 \pm 3.9$ | $4.9 \pm 4.0$ | $8.3 \pm 3.8$ |
| Banknote | $1.4 \pm 2.3$ | $1.1 \pm 1.1$ | $10.9 \pm 2.5$ | $10.3 \pm 2.9$ |
| Breastcancer | $2.5 \pm 0.7$ | $5.3 \pm 4.6$ | $7.2 \pm 2.0$ | $9.1 \pm 2.7$ |
| CMC | $-0.2 \pm 0.7$ | $15.1 \pm 4.7$ | $3.5 \pm 3.0$ | $5.7 \pm 3.3$ |
| HTRU2 | $0.7 \pm 0.3$ | $0.7 \pm 1.3$ | $9.2 \pm 3.1$ | $9.4 \pm 2.4$ |
| Phoneme | $3.5 \pm 2.9$ | $0.9 \pm 2.1$ | $6.8 \pm 0.7$ | $11.6 \pm 2.1$ |
| Ringnorm | $0.1 \pm 0.3$ | $1.7 \pm 0.5$ | $3.2 \pm 2.5$ | $6.4 \pm 2.9$ |
| Texture | $0.5 \pm 1.1$ | $1.2 \pm 0.8$ | $7.9 \pm 4.6$ | $4.9 \pm 3.9$ |
| Yeast | $-0.2 \pm 1.6$ | $1.9 \pm 3.8$ | $10.4 \pm 4.9$ | $2.3 \pm 4.6$ |

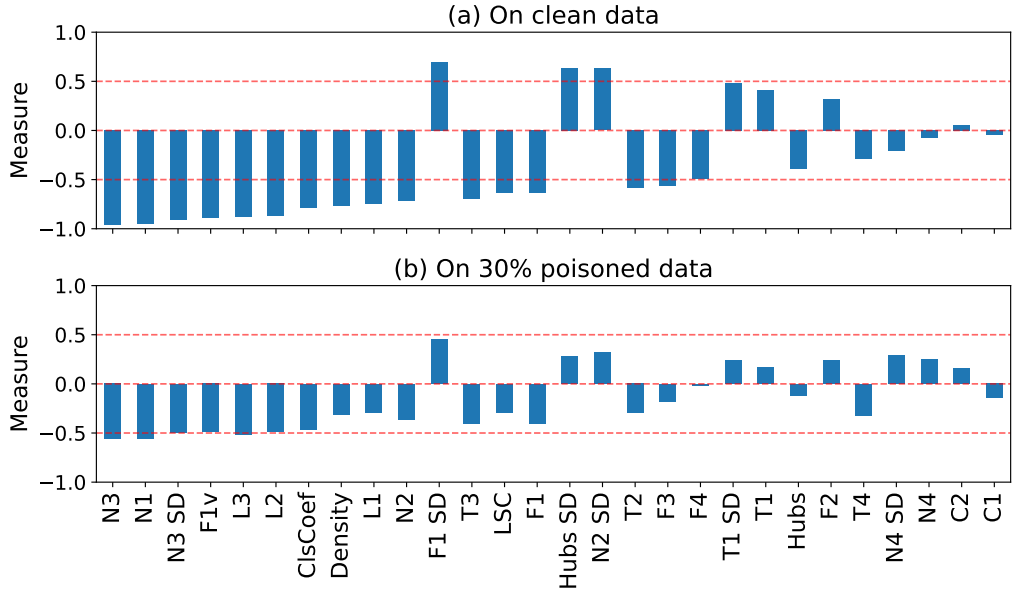

Figure 2: (a): Correlation of each measure in C-Measures to the test accuracy on synthetic datasets without poisoning attacks. (b): Same correlation but measured on the datasets containing 30% poisoning examples.

**Additional Results When Varying Thresholds.**

The results for DIVA with a single threshold set to 98% TNR are given in the paper. Here we provide additional results by varying the threshold. The first row in heatmaps represents the detectors' performance when there is no poison labels.

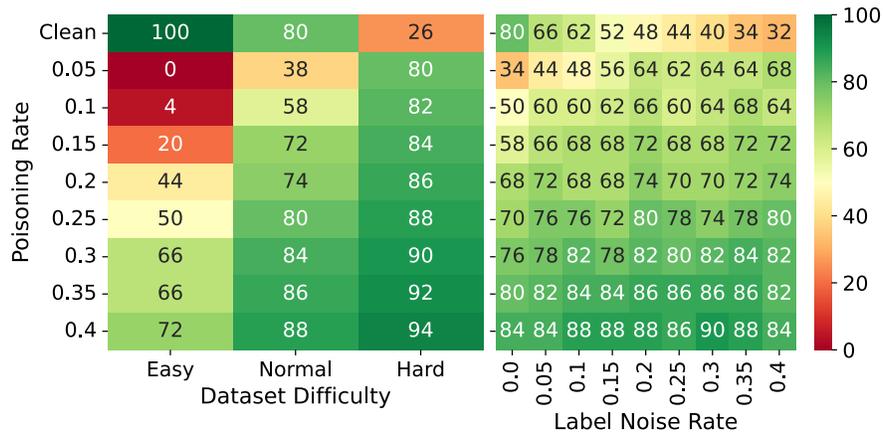

Figure 3: DIVA's performance with a fixed threshold set to 80% TNR.

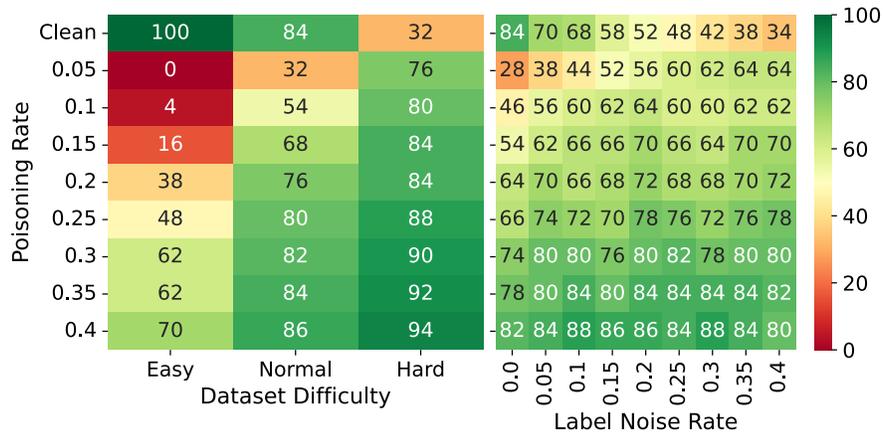

Figure 4: Diva's performance with a fixed threshold set to 85% TNR.

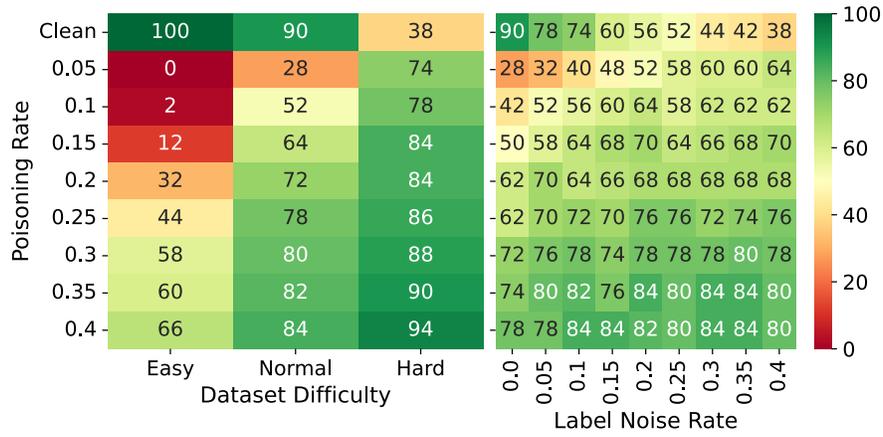

Figure 5: Diva's performance with a fixed threshold set to 90% TNR.

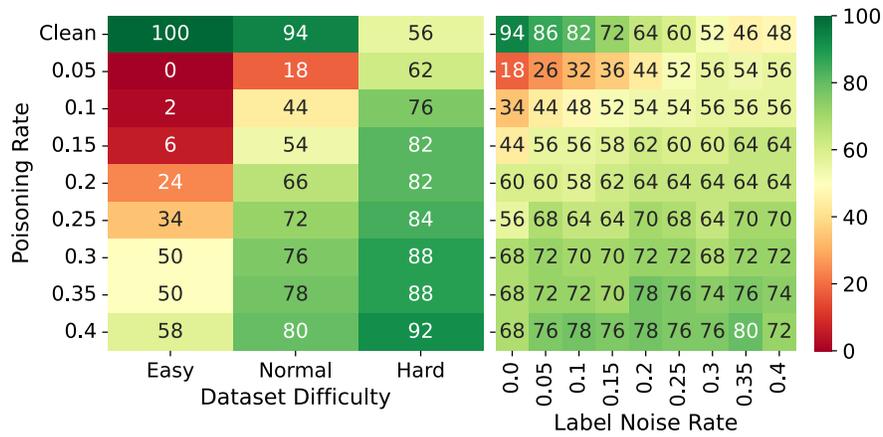

Figure 6: Diva's performance with a fixed threshold set to 95% TNR.

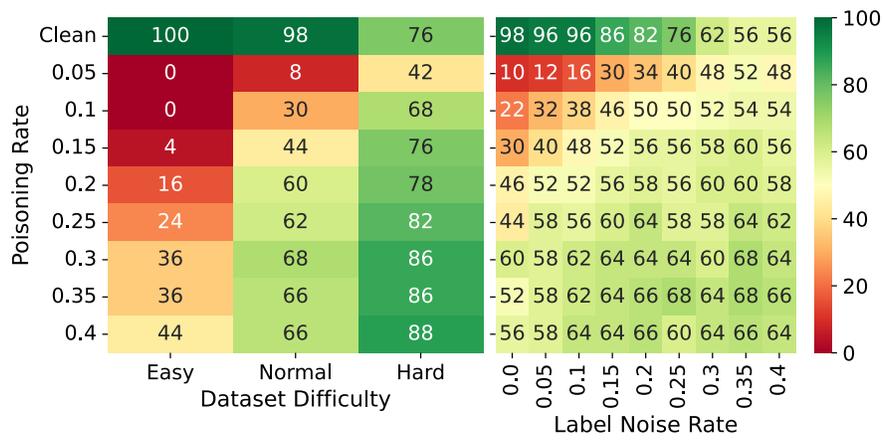

Figure 7: Diva's performance with a fixed threshold set to 99% TNR.



\end{document}

%% file: tikz/diva.tikz
\begin{tikzpicture}%[ultra thick]
    \newcommand \database [3]{
        \begin{tikzpicture}
            \draw[fill=#3] (0, #2) circle [x radius=#1, y radius=0.3*#1];
            % \draw (-#1, 0.5*#2) arc [start angle=180, end angle=360, x radius=#1, y radius=0.3*#1];
            \draw (-#1, 0*#2) arc [start angle=180, end angle=360, x radius=#1, y radius=0.3*#1];
            \draw (-#1,#2) -- (-#1,0) arc [start angle=180, end angle=360, x radius=#1, y radius=0.3*#1] -- (#1,#2);
        \end{tikzpicture}
    }
    
    \pgfdeclarelayer{bg}
    \pgfdeclarelayer{main}
    \pgfsetlayers{bg,main}
    \tikzstyle{block} = [rectangle, fill=white, align=center, minimum height=3em, text width=10em, draw, line width=0.1em]

    \newcommand \nodedist {1cm};      % distance of nodes (within one step)
    \newcommand \updowndist {-0.2cm};      % distance of nodes (within one step)
    \newcommand \divacol {gray!20};
    % \newcommand \divacol {my_yellow!30};

    % nodes
    \node [block, text width=6em] 
        (clf) {Train a\\classifier};
    \node [above right=\updowndist and \nodedist of clf, block]
        (accpoi) {evaluate empirical accuracy};
    \node [below right=\updowndist and \nodedist of clf, block]
        (acccl) {estimate clean accuracy\\(via meta-learner)};
    \node [draw=none] (helper) at (accpoi.east |- clf) {};
    % \node [right=\nodedist of helper.center, block]
    %     (diff) {calculate\\accuracy\\difference};
    % \node [right=\nodedist of diff, block] 
    %     (lr) {Threshold};
    \node [right=\nodedist of helper.center, block, text width=6em]
        (diff) {threshold\\accuracy\\difference};
    
    \node [above=3.2cm of $(clf)!0.5!(diff)$, block, draw, fill=\divacol] 
        (diva) {\diva};
    % \node [left=2*\nodedist of diva, cylinder, draw, shape border rotate=90, aspect=0.3, align=center, line width=0.1em] 
    %     (data) {Potentially\\poisoned\\input data};
    \node [left=2*\nodedist of diva, align=center] 
        (data) {\database{1.3cm}{1.5cm}{luke_red!30}};
    \node [below=-0.35cm of data.center, draw=none, fill=none, align=center] {Potentially\\poisoned\\input data};
    \node [right=2*\nodedist of diva] 
        (poison) {\includegraphics[width=1.5cm]{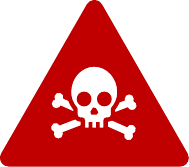}};

    % arrows
    \tikzstyle{every path}=[thick, rounded corners=5pt, black, -Stealth, shorten >= 2pt, shorten <= 2pt];
    \draw[shorten <= 0pt] (data) -- (diva);
    \draw[] (clf.east) -- ++(0.5*\nodedist,0) |- (accpoi);
    \draw[] (clf.east) -- ++(0.5*\nodedist,0) |- (acccl);
    \draw[] (accpoi.east) -- ++(0.5*\nodedist,0) |- (diff);
    \draw[] (acccl.east) -- ++(0.5*\nodedist,0) |- (diff);
    % \draw[] (diff) -- (lr);
    \draw[shorten >= 0pt] (diva) -- (poison);

    % background
    \begin{pgfonlayer}{bg}
        \node[rectangle, fill=\divacol, draw=none, fit=(clf)(accpoi)(acccl)(diff), inner xsep=0.5*\nodedist, inner ysep=0.2*\nodedist] 
            (R) {};
    \end{pgfonlayer}

    \node[below=0.2cm of diva.220, draw=none, fill=none] (h1) {};
    \node[below=0.2cm of diva.320, draw=none, fill=none] (h2) {};
    \draw[] (diva.220) -- (h1.center) -- (R.north west) |- (clf) {};
    \draw[] (diff) -| (R.north east) -- (h2.center) -- (diva.320) {};

\end{tikzpicture}

%% file: tikz/metabase_vertical.tikz
\begin{tikzpicture}
    \pgfdeclarelayer{bg}    % declare background layer
    \pgfsetlayers{bg,main}  % set the order of the layers (main is the standard layer)

    % INPUT: radius, height
    \newcommand \database [3]{
        \begin{tikzpicture}
            \draw[fill=#3] (0, #2) circle [x radius=#1, y radius=0.3*#1];
            % \draw (-#1, 0.5*#2) arc [start angle=180, end angle=360, x radius=#1, y radius=0.3*#1];
            \draw (-#1, 0*#2) arc [start angle=180, end angle=360, x radius=#1, y radius=0.3*#1];
            \draw (-#1,#2) -- (-#1,0) arc [start angle=180, end angle=360, x radius=#1, y radius=0.3*#1] -- (#1,#2);
        \end{tikzpicture}
    }
    % INPUT: radius, height
    \newcommand \metabase [2]{
        \begin{tikzpicture}
            \draw (0, #2) circle [x radius=#1, y radius=0.3*#1];
            \draw (-#1, 0.5*#2) arc [start angle=180, end angle=360,
                    x radius=#1, y radius=0.3*#1];
            \draw (-#1, 0.375*#2) arc [start angle=180, end angle=360,
                    x radius=#1, y radius=0.3*#1];
            \draw (-#1, 0) arc [start angle=180, end angle=360, x radius=#1, y radius=0.3*#1];
            \draw (-#1,#2) -- (-#1,0) arc [start angle=180, end angle=360,
                    x radius=#1, y radius=0.3*#1] -- (#1,#2);
            \path[draw=black, fill=black!15]
            (-#1, 0.5*#2) arc [start angle=180, end angle=360, x radius=#1, y radius=0.3*#1]
            -- (#1, 0.375*#2) arc [start angle=360, end angle=180,
                    x radius=#1, y radius=0.3*#1]
            -- (-#1, 0.5*#2);
        \end{tikzpicture}
    }

    % PARAM CONTROL
    \newcommand \cleancolor {luke_green!30};
    \newcommand \poisoncolor {luke_red!30};
    \newcommand \nodedist {1.5cm};      % distance of nodes (within one step)
    \newcommand \stepdist {1.6cm};        % distance of nodes (between steps)
    \newcommand \rowdist {1.5cm};       % distance of one row and nodes above
    \newcommand \datadist {1.6cm};        % distance of big dataset and nodes below
    \newcommand \dhw {0.5cm};           % data icon height and width
    \newcommand \rowfontsize {\small};  % font size for row

    % nodes
    \node[label={[align=center]below:Training\\Set}] (tr) {\database{\dhw}{\dhw}{\cleancolor}};
    \node[right = \nodedist of tr] (poi) {\database{\dhw}{\dhw}{\poisoncolor}};
    \node[right = 0.8*\nodedist of poi, rectangle, draw=black, outer sep=4, minimum width=1cm, minimum height=0.8cm] (clf) {Classifier};
    \node[right = \nodedist of clf, label={[align=center]below:Testing\\Set}] (te) {\database{\dhw}{\dhw}{\cleancolor}};
    \node[above = \datadist of $(tr)!0.5!(te)$, label=below:Dataset] (data) {\database{0.8}{0.8}{\cleancolor}};

    % row
    \node[below=\rowdist of poi.center] (r1) {{\rowfontsize 28 C-Measures}};
    \node[below=1.04*\rowdist of $(clf.east)!0.5!(te.west)$] (r3) {{\rowfontsize $\text{acc}_\text{clean}$}};
    \begin{pgfonlayer}{bg}
        \node[rectangle, fill=black!15, draw=black, fit=(r1)(r3)] (row) {};
    \end{pgfonlayer}

    % meta base and estimator
    \node[below=0.2cm of row.south west, anchor=north west, draw=none, fill=none] (metahelper) {};
    \node[right=0.4cm of metahelper.south, anchor=north west, label={[align=center]below:Meta-\\Database}] (meta) {\metabase{1}{2}};
    \node[right=1.5*\stepdist of meta.south east, anchor=south west, draw=none, label={[align=center]below:Meta-\\Learner}] (helper) {};
    \node[rectangle, draw=black, minimum width=1cm, minimum height=0.8cm, outer sep=4] (m) at (meta.190 -| helper) {m};

    % arrows in first step
    \tikzstyle{every path}=[thick, rounded corners=5pt];
    \draw[-Stealth] (data) -- (tr) {};
    \draw[-Stealth] (data) -- (te) {};
    \draw[-Stealth] (tr) -- (poi) node[midway,above] {Poison};
    \draw[-Stealth] (poi) -- node[midway,above] {Train} (clf) ;
    \draw[dotted, -Stealth] (clf) -- node[midway,above] (evalte) {Evaluate} (te);

    % arrows to row and meta
    \draw[-Stealth, shorten >=0.2cm] (poi) -- (r1) {};
    \draw[-Stealth, shorten >=0.3cm] (evalte)++(0,-0.4cm) -- (r3) {};
    \draw[-Stealth] (row.south) |- (metahelper.center) |- (meta.190) {};
    \draw[-Stealth] (meta.east |- m) -- node[midway,above] {Train} (m.west);

\end{tikzpicture}